\pdfoutput=1

\documentclass[11pt]{article}

\usepackage[final]{acl}

\usepackage{times}
\usepackage{latexsym}

\usepackage[T1]{fontenc}

\usepackage[utf8]{inputenc}

\usepackage{microtype}

\usepackage{inconsolata}

\usepackage{graphicx}
\usepackage{enumitem}
\usepackage{multirow}
\usepackage{amsmath}  
\usepackage{algorithm}  
\usepackage{algpseudocode}  

\usepackage{listings}
\usepackage{xcolor}
\usepackage{graphicx}
\usepackage{adjustbox}

\lstdefinestyle{python}{
	language=Python,
	basicstyle=\scriptsize\ttfamily,
	keywordstyle=\color{red}\ttfamily,
	stringstyle=\color{blue}\ttfamily,
	commentstyle=\color{gray}\ttfamily,
	morekeywords={},
	showstringspaces=false
}

\usepackage[hang, flushmargin]{footmisc}

\title{
PromptReps: Prompting Large Language Models to Generate Dense and Sparse Representations for Zero-Shot Document Retrieval
}

 \author{Shengyao Zhuang$^{1,2}$, Xueguang Ma$^{3}$, Bevan Koopman$^{1,2}$, Jimmy Lin$^{3}$, Guido Zuccon$^2$ \\ [1ex]
         $^1$CSIRO, \\ $^2$ The University of Queensland,
         \\ $^3$ University of Waterloo}
     
\begin{document}
\maketitle
\begin{abstract}

Utilizing large language models (LLMs) for zero-shot document ranking is done in one of two ways: (1) prompt-based re-ranking methods, which require no further training but are only feasible for re-ranking a handful of candidate documents due to computational costs; and (2) unsupervised contrastive trained dense retrieval methods, which can retrieve relevant documents from the entire corpus but require a large amount of paired text data for contrastive training. 
In this paper, we propose PromptReps, which combines the advantages of both categories: no need for training and the ability to retrieve from the whole corpus. Our method only requires prompts to guide an LLM to generate query and document representations for effective document retrieval. Specifically, we prompt the LLMs to represent a given text using a single word, and then use the last token's hidden states and the corresponding logits associated with the prediction of the next token to construct a hybrid document retrieval system. The retrieval system harnesses both dense text embedding and sparse bag-of-words representations given by the LLM. 
Our experimental evaluation on the MSMARCO, TREC deep learning and BEIR zero-shot document retrieval datasets illustrates that this simple prompt-based LLM retrieval method can achieve a similar or higher retrieval effectiveness than state-of-the-art LLM embedding methods that are trained with large amounts of unsupervised data, especially when using a larger LLM.

\end{abstract}

\begin{figure}
	\centering
	\includegraphics[width=\columnwidth]{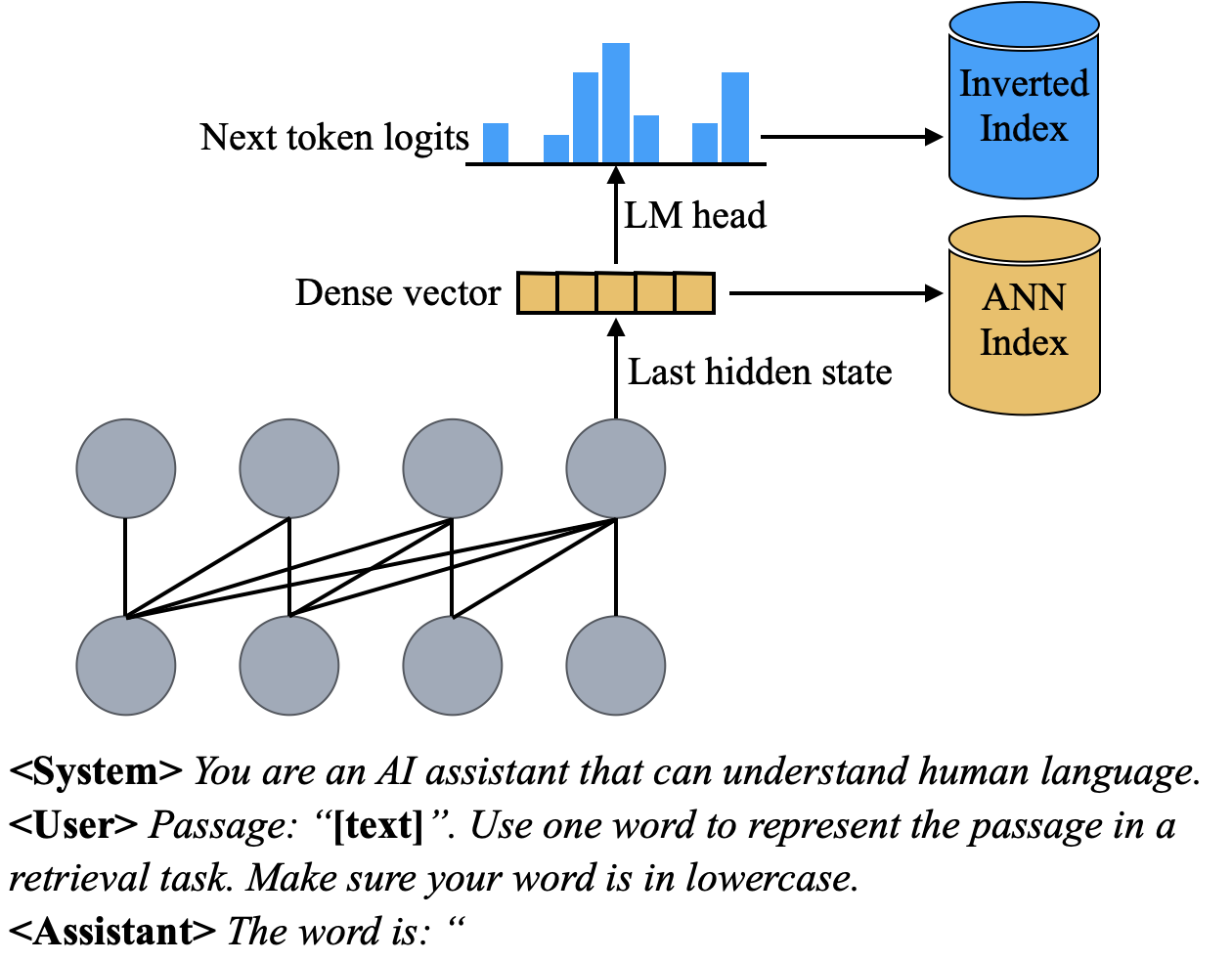}\vspace{-8pt}
	\caption{Overview of PromptReps. LLMs are prompted to simultaneously generate dense and sparse representations,  then used to build search indexes.}
	\label{fig:architecture}\vspace{-15pt}
\end{figure}

\section{Introduction}
Large Language Models (LLMs) such as GPT4 and LLaMA, which are pretrained on massive corpora and finetuned to follow user instructions, have strong zero-shot natural language understanding capabilities~\cite{openai2023gpt4, llama2}.
Via prompting, LLMs excel in various text generation tasks such as question answering, writing assistance, and conversational agent~\cite{hendryckstest2021, Liu_2023}.
Inspired by the success of LLMs on natural language understanding tasks, research has explored the potential of using LLMs to perform unsupervised document ranking.

One line of work focuses on directly prompting LLMs to infer document relevance to a given query~\cite{sachan-etal-2022-improving,zhuang-etal-2023-open,ma2023zeroshot,Sun2023IsCG,pradeep2023rankvicuna,zhuang2023setwise,qin2024large}. For instance, RankGPT~\cite{Sun2023IsCG} casts document re-ranking as a permutation generation task, prompting LLMs to generate re-ordered document identifiers according to the document's relevance to the query. These methods leverage LLMs for document ranking in a complete zero-shot setting where no further training is required. However, these methods can only serve as a second-stage re-ranker on a handful of candidate documents. This is because each prompt requires one full LLM inference: for example, in the case of a corpus with 1M documents, a pointwise approach would require the construction of 1M prompts and thus the execution of 1M (costly) LLM inferences -- making it unfeasible for an online search engine. 

Another line of research leverages LLMs as a text embedding model for dense document retrieval~\cite{lee2024gecko,wang2024text,wang2024improving,behnamghader2024llm2vec}. For example, E5-mistral~\cite{wang2024improving} employs LLMs to create synthetic datasets of query-document pairs. These paired text data are then used to perform unsupervised contrastive training for a Mistral LLM-based dense retriever. Since the queries and documents are encoded with LLMs separately; i.e., using a bi-encoder architecture, these methods could serve as a first-stage document retriever. However, all existing LLM-based retrievers require an unsupervised contrastive training step to transform a generative LLM into a text-embedding model. Even with parameter-efficient training techniques such as LoRA~\cite{hu2022lora}, this extra training is still very expensive. 
For example, the contrastive training of E5-mistral using a large batch size (2048) and LoRA took $\approx18$ hours on 32 V100 GPUs~\cite{wang2024improving}. 
 
In this work, we propose a new zero-shot LLM-based document retrieval method called PromptReps.
We demonstrate that LLMs can be directly prompted to produce query and document embeddings, which can serve as effective text representations for neural retrieval systems.
Specifically, we prompt an LLM by asking it to use a single word to represent a query or a document.
Then, we extract the last layer's hidden state of the last token in the prompt as the dense representation of the input text.
Simultaneously, we utilize the logits associated with predicting the subsequent token to form a sparse representation.
As illustrated in Figure~\ref{fig:architecture}, through a single forward pass, we generate text representations for a document that can be indexed for dense, sparse, or hybrid search architectures. We also explore alternative representations in addition to the core idea in this paper, where we generate multiple words, and use multiple embeddings to represent an item (Figures~\ref{fig:single_rep} and~\ref{fig:colbert_rep}).


Our empirical evaluation on multiple datasets show that PromptReps can achieve a similar or higher zero-shot retrieval effectiveness than previous trained LLM-based embedding methods, especially when a large LLM is utilized. Of key importance is that our method is the first LLM-based method that can effectively perform full corpus retrieval while at the same time not requiring contrastive training, demonstrating that prompt engineering for generative LLMs is capable of generating robust representations for retrieval. 

Code for fully reproducing our results is available at \url{https://github.com/ielab/PromptReps}.


\section{Related Work}
\subsection{Supervised neural retrievers}
Neural retrievers based on the bi-encoder architecture bring significant improvements over traditional best-match retrievers such as BM25. Dense retrievers such as DPR~\cite{dpr} and ANCE~\cite{ance} are based on transformer language models and encode text into low-dimensional vectors, conducting search with nearest neighbor search. On the other hand, sparse neural retrievers such as DeepImpact~\cite{deepimpact}, uniCOIL~\cite{lin2021brief}, TILDE~\cite{tilde,zhuang2021fast}, and SPLADE~\cite{splade}, also based on transformer language models, encode text into high-dimensional sparse vectors as bag-of-words representations, conducting search with inverted index. Recent works also explored fine-tuning generative LLMs as dense retrievers such as RepLLaMA~\cite{ma2023finetuning} and LLaRA~\cite{liao2024llara}. A hybrid neural retrieval system refers to a system that combines the rankings provided by both dense and sparse retrievers, often resulting in an enhanced final ranking~\cite {lin2021brief,bm25interpolation}. 

All these retrievers are trained with supervised relevance judgment data (e.g., MSMARCO~\cite{bajaj2018ms}) using contrastive learning. Our work instead focuses on building a hybrid neural retrieval system with zero-shot dense and sparse document representations without supervised contrastive learning and based on generative LLMs. This capability has two implications: (1) no contrastive training is required, which is expensive when applied to LLMs with several billions parameters, and (2) no human-labelled training data is required, which may be laborious and expensive to obtain. With regards to the first point, \citet{wang2024improving} reported that the training of E5-mistrail (7B parameters) took about 18 hours on 32 V100 GPUs, for an approximate cost of USD \$2,300,\footnote{Based on 4 On-Demand p3dn.24xlarge instances, June 2024.} emissions of $\approx$5.6 kgCO2e and consumption of $\approx$37.7 L of water for the associated cooling activities.\footnote{Emissions and water consumption estimates obtained using the frameworks of \citet{scells2022reduce,zuccon2023beyond}.} Scaling this training to more and larger LLMs, and more data, will consequently further increase costs. Our proposed method does not incur these additional contrastive pre-training costs. 
With regards to the second point, dense retrievers have shown to have poor generalisability when applied to data out-of-domain or out-of-task compared to the data used for contrastive training~\cite{thakur2021beir,zhuang2021dealing,zhuang2022char,ren-etal-2023-thorough,lin2023train,Lupart2022MSShiftAA}. In presence of shift in data between training and deployment, retrieval losses can be significant: dense retrieval effectiveness can plummet far below that of best-match models like BM25~\cite{khramtsova2023selecting,khramtsova2024leveraging}. The acquisition of in-domain/in-task training data can be costly, laborious and impractical/impossible especially in domain-specific applications when dealing with sensitive, private data.  


\subsection{Unsupervised neural retrievers}
There have also been attempts at training effective neural retrievers without relying on human relevance judgments.
Methods such as Contriever~\cite{izacard2022unsupervised} and E5~\cite{wang2024text}, train a dense retriever with large-scale pseudo query-document pairs to build unsupervised training data.
LLMs have also been adapted as unsupervised text embedding models for first-stage document retrieval.
For instance, HyDE~\cite{gao-etal-2023-precise} enhances query representations for an unsupervised retriever by replacing the original query with LLM-generated hypothetical documents. 

More recent work has focused on directly converting generative LLMs into a text-embedding model with unsupervised contrastive pre-training. Methods like E5-Mistral-Inst~\cite{wang2024improving} and Gecko~\cite{lee2024gecko} use large-scale weakly supervised paired text data or LLM-generated query-document pair data to perform contrastive training on top of LLMs. 
LLM2Vec~\cite{behnamghader2024llm2vec}, on the other hand, conducts further masked next token prediction pre-training with bidirectional attention, and SimCSE~\cite{gao-etal-2021-simcse} trains on raw text data to transform LLMs into text encoders. Although no labeled data is used, these methods require synthetic or unsupervised paired text data to perform contrastive pre-training (thus still experiencing training costs in terms of computations; and further computational costs may be associated with the generation of synthetic training data). 
Our method instead relies solely on prompt engineering to transform LLM into a robust text encoder for document retrieval without any extra training.

\subsection{Prompting LLMs for document ranking}
Inspired by the prompt-following capacity of LLMs, recent studies have explored prompting LLMs for document re-ranking. For instance, UPR~\cite{sachan-etal-2022-improving} ranks documents pointwise by prompting the LLM to generate a relevant query for a given document and rank documents based on the likelihood of generating the query. RankGPT~\cite{Sun2023IsCG} and LRL~\cite{ma2023zeroshot} propose to re-rank a list of documents at once and generate permutations for the reordered list. Pairwise~\cite{qin2024large} and Setwise~\cite{zhuang2023setwise} prompting methods have also been explored to improve effectiveness and efficiency in the LLM re-ranking pipeline. These methods are only feasible for re-ranking a handful of candidate documents, thus limited to second-stage document re-ranking. In contrast, our approach utilizes prompts to construct the first-stage retrievers.

\subsection{Prompting LLMs for sentence embedding}
The methods most similar to ours prompt LLMs to generate sentence embeddings for semantic textual similarity (STS) tasks~\cite{jiang2023scaling,lei2024metatask,zhang2024simple}. These previous methods also used an Explicit One-word Limitation (EOL) prompt, which also instructs LLMs to represent a sentence with one word. However, these methods only evaluate such prompts on STS datasets, and their effectiveness on information retrieval datasets with large document corpora is unknown. Additionally, these methods only represent text with dense embeddings from the hidden states; our method instead generates dense and sparse representations simultaneously to build a hybrid retrieval system. Our empirical results show that dense embeddings alone perform poorly for document retrieval tasks with some LLMs, but sparse representations are much more robust, and the best retrieval effectiveness is achieved with the hybrid retrieval system with scaled model size.

%
%
%

\section{PromptReps}
Previous work that leverages LLMs for document ranking are limited to document re-ranking tasks with prompts or rely on contrastive learning to transform a generative LLM into an embedding model for document retrieval. Unlike these previous works, here we aim to directly prompt LLMs to generate both dense embedding representations and sparse bag-of-words representations for document retrieval without any form of extra training effort. To achieve this, we devise the prompt as illustrated in Figure~\ref{fig:architecture} as the input text for LLMs, where \textbf{<System>} \textbf{<User>} and \textbf{<Assistant>} are LLM pre-defined conversational prefix tokens and \textbf{[text]} is the placeholder for passage text. 

%
%

When using this prompt for text generation, the language model needs to find a single word in its token vocabulary that can best represent the given passage to generate. However, since there could be multiple words to represent the passage, there might be multiple tokens in the vocabulary that have a high probability of being sampled by the language model. Such a distribution over the vocabulary, which is often refers to as
``logits'', could provide a good representation of the given passage. In addition, since the logits are computed by the last layer hidden state\footnote{Often through dot product between the last hidden state with all token embeddings.} of the last input token (` \textbf{``} '), which is a dense vector embedding, it could also serve as a dense representation of the passage.

Based on the above intuition, we develop a sparse + dense hybrid document retrieval system by utilizing both the next token logits and the last layer hidden states outputted by the LLM with our designed prompt. 

Specifically, during the document indexing phase, we pass all the documents (one at the time) with our prompt into the LLM to get output hidden states and logits. To build a sparse retrieval pipeline with logits, we first need to sparsify the logits representation to be able to perform efficient sparse retrieval. This is because logits originally had values for all tokens in the vocabulary, essentially forming dense vectors with dimensions equal to the vocabulary size. To sparsify the logit representations for sparse retrieval, we perform the following steps: 
\begin{table*}[]
	\centering
	\caption{nDCG@10 scores of BEIR 13 publicly available datasets.}\vspace{-5pt}
	\resizebox{\textwidth}{!}{
		\begin{tabular}{l|c|cc|cc|c c c|c c c  }
			\hline
			\multicolumn{2}{c|}{}&  \multicolumn{2}{c|}{Sup Contrastive training } &\multicolumn{2}{c|}{Unsup Contrastive training } & \multicolumn{6}{c}{PromptReps (ours)}  \\\hline 
			\multicolumn{1}{r|}{LLM} &-&BERT110M&BERT110M&BERT330M & Llama3-8B-I & \multicolumn{3}{c|}{Llama3-8B-I} & \multicolumn{3}{c}{Llama3-70B-I} \\
			Dataset & BM25 &SPLADE++&DRAGON+& E5-PT$_\textrm{large}$ & LLM2Vec & Dense & Sparse & Hybrid & Dense & Sparse & Hybrid\\
			\hline
			arguana & 39.70 &52.1 & 46.9& 44.4 & 51.73& 29.70 & 22.85 & 33.32 & 31.65 & 24.66 & 35.27 \\
			climatefever & 16.51 &22.8&22.7 & 15.7 & 23.58 & 19.92 & 9.98 & 21.38 & 19.95 & 12.14 & 22.18  \\
			dbpedia & 31.80 &44.2& 41.7 & 37.1 & 26.78 & 31.53 & 28.84 & 37.71 & 31.12 & 28.30 & 37.59 \\
			fever & 65.13 &79.6&78.1& 68.6 & 53.42 & 56.28 & 52.35 & 71.11 & 42.06 & 51.75 & 63.97 \\
			fiqa & 23.61 &35.1&35.6& 43.3& 28.56 & 27.11 & 20.33 & 32.40 & 30.80 & 22.16 & 34.66 \\
			hotpotqa & 63.30 &68.6 &66.2&52.2 & 52.37 & 19.64 & 44.75 & 47.05 & 24.32 & 42.12 & 48.51 \\
			nfcorpus & 32.18 &34.5&33.9& 33.7 & 26.28 & 29.56 & 28.18 & 32.98 & 33.84 & 29.74 & 36.08  \\
			nq & 30.55 &54.4&53.7& 41.7 &  37.65 & 34.43 & 29.55 & 43.14 & 38.25 & 30.37 & 46.97 \\
			quora & 78.86 &81.4&87.5& 86.1& 84.64 & 81.77 & 70.35 & 84.24 & 81.18 & 67.69 & 83.70 \\
			scidocs & 14.90 &15.9&15.9& 21.9& 10.39 & 18.51 & 11.57 & 17.59 & 20.59 & 13.25 & 19.10  \\
			scifact & 67.89 &69.9&67.9& 72.3 & 66.36 & 52.68 & 58.48 & 65.71 & 63.12 & 61.53 & 70.34 \\
			trec-covid & 59.47 &71.1& 75.9& 62.1 & 63.34 & 59.52 & 54.59 & 69.25 & 67.64 & 63.00 & 76.85 \\
			touche & 44.22 &24.4&26.3& 19.8 & 12.82 & 14.85 & 18.47 & 21.65 & 15.56 & 18.65 & 22.35 \\\hline
			avg & 43.70 &50.3&50.2& 46.06 & 41.38 & 36.58 & 34.64& 44.43 & 38.47 & 35.80 & 45.97 \\
			\hline
		\end{tabular}
	}\label{tab:results}\vspace{-10pt}
\end{table*}

\begin{enumerate}[leftmargin=5mm,itemsep=0.00em]
    \item Lowercase the input document text to align with the phrase ``Make sure your word is in lowercase.'' in the prompt since this phrase skewed the sampling distribution towards lowercase tokens (a ``sparser'' distribution). We then utilize the NLTK toolkit~\cite{bird-loper-2004-nltk} to extract all words in the document, filtering out standard English stopwords and punctuation.
    \item Next, we use the LLM's tokenizer to tokenize each extracted word and obtain their token IDs.\footnote{Note that many words may be split into sub-tokens, resulting in multiple token IDs, all of which are considered in the logits} We retain only the values corresponding to the obtained token IDs in the logits and set the rest of the dimensions to zero, thereby considering only tokens present in the documents, thus enabling exact term matching in retrieval. 
    \item Next, we follow the SPLADE recipe~\cite{splade}, using the ReLU function to remove dimensions with negative values and applying log-saturation to the logits to prevent certain tokens from dominating. To further enhance the sparsity of logits, we only keep tokens within the top 128 values if the logits had more than 128 non-zero values after the previous steps.
    \item Finally, the logits are quantized by multiplying the original values by 100 and taking the integer operation on that, and the obtained values represent the weights of corresponding tokens.
\end{enumerate}\vspace{-5pt}
With these adjustments, the logits representations of documents are heavily sparsified, allowing for efficient sparse retrieval with an inverted index.

For dense retrieval, we directly use the hidden states as the embeddings of the documents. For indexing these embeddings, we simply normalize all the embeddings and add them into an Approximate Nearest search (ANN) vector index. In Appendix~\ref{appendix:python}, we provide example Python code of generating dense and sparse representations with PromptReps.

At query time, we process the queries exactly the same as the documents, with the only exception being that the term ``passage'' in the prompt is replaced with ``query''.\footnote{The only exception in our experiments is the Quora dataset, which is a duplicate query search task. Therefore, we use the query prompt for both queries and documents.} The dense representation of the query is utilized for semantic search via the ANN index, while the sparse representation of the query is employed for exact term matching via the inverted index. Following previous work~\cite{bm25interpolation}, we compute the final document scores by applying min-max normalization to both dense and sparse document scores. These normalized scores are then linearly interpolated with equal weights to produce the final document scores. We do not explicitly tune the weight because our setting is zero-shot retrieval, and we wanted to maintain the ``zero-shot'' nature of our approach. Nevertheless, in Appendix~\ref{appendix:weights}, we explore the impact of the different weight settings.


\section{Experimental Setup}
\textbf{Dataset and evaluation:} We evaluate the document ranking effectiveness of both baseline methods and our proposed PromptReps using MSMARCO~\cite{bajaj2018ms} passage retrieval, TREC deep learning~\cite{craswell2020overview} and BEIR~\cite{thakur2021beir}. These datasets encompass various IR tasks, providing a heterogeneous evaluation environment. For MSMARCO we report MRR@10 and for TREC deep learning and BIER we report nDCG@10 scores, the commonly employed evaluation measure for these datasets.

\noindent\textbf{Baselines:} We compare PromptReps with strong unsupervised first-stage retrievers including BM25, a classic term frequency-based sparse retrieval method, and E5-PT$_\textrm{large}$~\cite{wang2024text}, a state-of-the-art BERT large-based dense embedding method trained on 1.3B carefully crafted unsupervised text pairs. LLM2Vec~\cite{behnamghader2024llm2vec}, a Llama3-8B-Instruct LLM-based dense embedding method trained with bi-directional attention, masked next token prediction, and SimCSE~\cite{gao-etal-2021-simcse} on the Wikipedia corpus.
In addition, We also report state-of-the-art supervised contrastive, fine-tuned BERT-based sparse retriever SPLADE++~\cite{splade++} and dense retriever DRAGON+~\cite{lin2023train}. We note that these methods are trained with lots of supervised training data and knowledge distillation from teacher models, thus it is unfair to compare with our method and other unsupervised baselines. However, we think it is useful to compare with supervised methods to understand the gap between supervised and unsupervised methods.

\noindent\textbf{Implementation of PromptReps:} PromptReps is implemented using four base LLMs: Mistral-7b-Instruct-v0.2\footnote{https://huggingface.co/mistralai/Mistral-7B-Instruct-v0.2}~\cite{jiang2023mistral}, Phi-3-mini-4k-instruct\footnote{https://huggingface.co/microsoft/Phi-3-mini-4k-instruct}~\cite{abdin2024phi3}, Llama3-8B-Instruct,\footnote{https://huggingface.co/meta-llama/Meta-Llama-3-8B-Instruct} and Llama3-70B-Instruct\footnote{https://huggingface.co/meta-llama/Meta-Llama-3-70B-Instruct}~\cite{llama3modelcard}. Dense and sparse document and query encodings are implemented using the Huggingface Transformers library~\cite{wolf-etal-2020-transformers} and the Tevatron toolkit~\cite{tevatron}. The Faiss library~\cite{douze2024faiss} is used to build the ANN index with cosine similarity as the embedding distance metric. We simply use brute force search for ANN (IndexFlatIP in Faiss) for a fair comparison with the baselines. For sparse retrieval, Pyserini~\cite{pyserini} is utilized to construct the inverted index. For the dense and sparse ranking hybrid, the Ranx library~\cite{ranx} is employed. In our experiments, we report dense only, sparse only, and the full hybrid results.

\vspace{-4pt}
\section{Zero-shot Results}
We start by showing our overall results on the BEIR dataset, which we treated as test set; we then analyse choices in instantiation of PromptReps, including different variations in the prompt using the MSMARCO and TREC deep learning datasets, which we used as development datasets to inform the choices we made to run PromptReps on BEIR.

\subsection{Zero-shot retrieval effectiveness on BEIR}

We present our results on BEIR in Table~\ref{tab:results}. 
The first observation highlights that BM25 is a strong zero-shot retrieval method, capable of outperforming LLM2Vec, based on the Llama3-8B-Instruct LLM, across numerous datasets, achieving a higher average nDCG@10 score. This outcome implies that even with a large-size LLM, bi-directional attention enabled, additional pre-training, and SimCSE-based unsupervised contrastive training, there remains a gap in transforming a decoder-only LLM into an effective retrieval method.

On the other hand, E5-PT$_\textrm{large}$, based on the BERT-large model, is the first method that can outperform BM25 without any supervised training data. However, it has been trained on a massive, carefully mined text pair dataset with a large batch size, which may require more data-collecting efforts and computational resources than LLM2Vec.

PromptReps with Llama3-8B-Instruct LLM has lower nDCG@10 scores when only using dense or sparse retrieval. However, the hybrid system (combining dense and sparse) contributes notable retrieval effectiveness improvements, surpassing BM25 and approaching the state-of-the-art E5-PT$_\textrm{large}$. Notably, this is achieved without any form of extra training but solely relying on prompts. 

The scaling law observed for LLMs~\cite{kaplan2020scaling} and dense retrievers~\cite{scalingDR} also applies here. When changing Llama3-8B-Instruction to Llama3-70B-Instruction, the dense and sparse retrieval effectiveness of PromptReps further improves, with the hybrid approach comparable to E5-PT$_\textrm{large}$. 
\begin{table}[]
	\centering
	\small
	\caption{nDCG@10 scores of unsupervised LLM-based methods hybrid with BM25 on BEIR 13 publicly available datasets. D+S+BM25 stands for PromptReps hybrid with its dense, sparse and BM25.}\vspace{-5pt}
	\resizebox{\columnwidth}{!}{
		\begin{tabular}{l|c|c|c }
			\hline
			\multicolumn{1}{r|}{LLM} & BERT-330M & Llama3-8B-I &Llama3-70B-I\\
			Dataset & E5-PT+BM25 & D+S+BM25 & D+S+BM25  \\
			\hline
			arguana & 47.34 &	38.13 &39.53 \\
			climatefever & 	21.92 & 22.95 & 23.34 \\
			dbpedia & 43.46 &	40.87 & 41.63 \\
			fever &	78.04 & 	77.07&74.06 \\
			fiqa & 42.88 &	34.11&35.35\\
			hotpotqa &	69.18 &64.38 & 65.29\\
			nfcorpus & 	36.61 &	35.20&37.64 \\
			nq & 47.71 &	45.12&48.30\\
			quora &88.63&	86.26&86.60\\
			scidocs &	20.76&	17.97&18.82  \\
			scifact &	76.37&	70.92& 73.58 \\
			trec-covid & 74.09&	76.17&80.29 \\
			touche &35.01&	29.13& 34.15 \\\hline
			avg & 	52.46 &	49.10& 50.66 \\
			\hline
		\end{tabular}
	}\label{tab:bm25results}\vspace{-10pt}
\end{table}


\subsection{Further hybrid with BM25}

\begin{table*}[!htp]
\vspace{-1pt}
\centering
	\caption{Investigated prompts. The systems prompt and any text string before the prompts in this table are the same as Figure~\ref{fig:architecture}, thus omitted. \textit{<A>} denotes the model-specific assistant special token.}\label{tab:prompts}\vspace{-5pt}
	\resizebox{\textwidth}{!}{
		\begin{tabular}{r|l}\hline
			ID& Prompts \\\hline
			1 & Use one word to represent the passage in a retrieval task.\textit{<A>}The word is: " \\
			2 & Use one word to represent the passage.\textit{<A>}The word is: " \\
			3 & Use one most important word to represent the passage in a retrieval task. Make sure your word is in lowercase.\textit{<A>}The word is: " \\
			4 & Use one word to represent the passage in a retrieval task.\textit{<A>}\\
			5 & Use one most important word to represent the passage in a retrieval task.\textit{<A>}The word is: "  \\
			6 & Use one word to represent the passage in a retrieval task. Make sure your word is in lowercase.\textit{<A>}The word is: "\\\hline
		\end{tabular}
	}\vspace{-10pt}
\end{table*}

In Table~\ref{tab:bm25results} we report results of unsupervised LLM-based retrievers further hybrid with BM25 on BEIR datasets. For PromptReps, we generate hybrid ranking by combining dense, sparse, and BM25 rankings using min-max normalization, assigning equal weight to each. For the baseline, we report E5-PT$_\textrm{large}$ hybrid with BM25, also using min-max normalization and equal weights. Compared to their standalone retrieval effectiveness reported in Table~\ref{tab:results}, the effectiveness of all LLM-based retrievers significantly improved. E5-PT$_\textrm{large}$ and PromptReps with Llama 70B model surpassed supervised training methods. These results demonstrate that it is possible to build a strong retrieval system with LLMs and BM25 without the need for any supervised training.
\vspace{-1pt}
\subsection{Sensitivity to different prompts}
In the previous experiments, we always use the prompt illustrated in Figure~\ref{fig:architecture}. In this section, we study how different prompts impact the retrieval effectiveness. Particularly, we design six different prompts,\footnote{The prompt in Figure~\ref{fig:architecture} is prompt  \#6 in Table~\ref{tab:prompts}.} listed in Table~\ref{tab:prompts}, and conduct experiments on TREC deep learning 2019 and 2020 datasets, and MSMARCO passage retrieval dev sub-dataset. 
We use Llama3-8B-Instruction as the base LLM for PromptReps. The results are listed in Table~\ref{tab:prompt_results}. We also report results of Recall@1000 and other base LLMs in Appendix~\ref{appendix:results}.

\begin{table}[h]\centering
\small
	\caption{Retrieval effectiveness of different prompts on TREC deep learning and MSMARCO. The ID correspond to the prompt IDs list in Table~\ref{tab:prompts}.}\label{tab:prompt_results}\vspace{-4pt}
	\resizebox{1\columnwidth}{!}{
		\begin{tabular}{r|c|c|c|c}\hline
			ID&Methods & DL2019 & DL2020 & MSMARCO\\\hline
			- &BM25 &49.73 &48.76  &18.75  \\
			- &LLM2Vec &- &- &13.61  \\\hline
			\multicolumn{5}{c}{PromptReps  Llama3-8B-Instruct (ours)} \\\hline
			1 &Dense &49.26  &40.28  &16.26  \\
			2 &Dense &43.32 &31.60  &12.52  \\
			3 &Dense &49.20 &\textbf{43.90} &17.49  \\
			4 &Dense &0.00  &0.00 &0.00  \\
			5 &Dense &47.19  &40.17  &16.02 \\
			6 &Dense &\textbf{50.62} &43.81 &\textbf{17.54} \\ \hline
			1 &Sparse &41.77  &44.81  &20.12  \\
			2 &Sparse &39.90  &43.10  &19.13  \\
			3 &Sparse &\textbf{43.50} &44.87  &20.42  \\
			4 &Sparse &21.77  &20.49  &7.22 \\
			5 &Sparse &42.18 &44.17  &19.78 \\
			6 &Sparse &42.25  &\textbf{45.60}  &\textbf{20.85} \\\hline
			1 &Hybrid &53.67  &54.35  &23.68  \\
			2 &Hybrid &50.65 &49.25  &21.76 \\
			3 &Hybrid &\textbf{55.64}  &53.83 &23.86  \\
			4 &Hybrid &13.47  &11.81 &5.06 \\
			5 &Hybrid &54.16  &52.06 &23.25  \\
			6 &Hybrid &55.58  &\textbf{56.66}  &\textbf{24.62} \\\hline
		\end{tabular}	
	}\vspace{-10pt}

\end{table}

The results demonstrate that PromptReps can achieve a similar level of retrieval effectiveness as BM25 and surpass LLM2Vec with most of the prompts. The only prompt that does not work well is prompt \#4, which does not include the phrase ``\textit{The word is: ``}'' to force the LLM to generate the representative word as the next token. This is expected because, without this phrase, the first generated token would be a general token such as ``\textit{The}'' which is not representative of the input text.

Interestingly, our results also show that LLMs have instruction-following ability in this representation generation task. For instance, comparing prompts \#1 and \#2, the only difference is the phrase ``\textit{in a retrieval task}'', and the prompt with this phrase yields higher retrieval effectiveness across all datasets. Additionally, comparing prompts \#1 and \#6, the difference is the phrase ``\textit{Make sure your word is in lowercase}'', which matches our sparse exact matching mechanism where we first lowercase the input text. This phrase can further improve the retrieval effectiveness. Finally, using the adjective phrase ``\textit{most important}'' in the prompt does not significantly impact the results.

\subsection{Impact of different LLMs}
In this section, we explore how different base LLMs impact PromptReps. For this study, we investigate five state-of-the-art open-sourced decoder-only LLMs, covering different model sizes and models with or without instruction tuning. We use prompt \#6 for all LLMs\footnote{Only model specific conversational special tokens are changed.} and report MRR@10 scores on the MSMARCO datasets. The results are illustrated in Figure~\ref{fig:different_llms}; more detailed results including on TREC deep learning datasets are reported in Appendix~\ref{appendix:results}.

The results show that the hybrid retrieval effectiveness of PromptReps consistently outperforms BM25, regardless of which LLM is used, with the only exception of Mistral-7B-Instruct. When using the Mistral-7B-Instruct LLM, the dense-only retriever performs poorly. Surprisingly, implementing PromptReps with Phi-3-mini-4k-instruct achieved much higher retrieval effectiveness than that of Mistral-7B-Instruct, despite having far fewer parameters (3.8B).

Meta-Llama-3 models are generally very effective for our method. For 8B models, the instruction-tuned model performs significantly better than the pretrained-only model, indicating that the instruction fine-tuning is helpful to further improve our method. The 70B instruction-tuned model achieved the best hybrid retrieval results, but the dense-only and sparse-only retrieval effectiveness is similar to the 8B instruction-tuned model. These results agree with the BEIR results presented in Table~\ref{tab:results}.

\begin{figure}
	\centering
	\includegraphics[width=1\columnwidth]{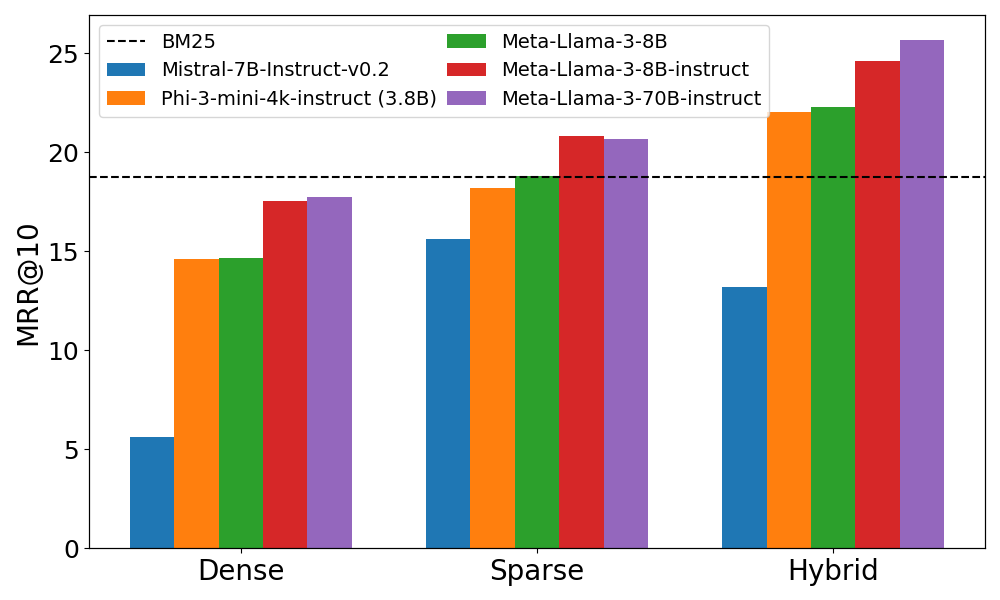}
	\caption{MRR@10 scores on MSMARCO of PromptReps with different LLMs.}
	\label{fig:different_llms}
\end{figure}

\section{Supervised Results}\label{sec:supervised}
We have demonstrated the strong zero-shot effectiveness of PromptReps. Now we explore the question: Can PromptReps serve as a better initialization for LLM-based embedding models in downstream contrastive supervised fine-tuning? 

To address this question, we conduct supervised fine-tuning experiments using MSMARCO. Specifically, we follow the RepLlama training recipe~\cite{ma2023finetuning} to fine-tune the LLama-3-8B-Instruct base model with InfoNCE loss and hard negative passages mined by a BM25 and dense retrieval hybrid system. The detailed training hyper-parameters are listed in Appendix~\ref{appendix:parameters}. For the RepLlama baseline, we adhere to the original implementation, which appends the prefixes ``Query: '' and ``Passage: '' to the query and document text, respectively, and adds the end-of-sentence token at the end of the text. The output embedding of this token is then used to represent the text. For PromptReps, we use our proposed prompt (\#6 in Table~\ref{tab:prompts}) and the last token hidden states and logits as the dense and sparse representation of the text. For fine-tuning PromptReps, we explore two settings, \textit{PromptReps-dense only}, which only uses the dense representation of PromptReps to calculate document scores during training and inference. This setting ensures a fair comparison with RepLlama, as the only difference is the prompt used. The other setting involves using both dense and sparse document scores to calculate the loss, simply adding the two losses as the final loss. During inference, we report the dense, sparse, and hybrid retrieval effectiveness separately for this setting.

We train both RepLlama and PromptReps for 1 epoch using the full MSMARCO training data, which contains 490k training examples. In addition to the full training, we also explore a low-resource training setting, where we sample 1k examples from the entire MSMARCO dataset. We then split the 1k examples into a training and a validation set with a 9:1 ratio. We monitor the validation loss after each training epoch and stop the training, selecting the best checkpoint if no lower validation loss is observed for three consecutive epochs.

The results are shown in Table~\ref{tab:supervise}. Surprisingly, with only 1k training examples, ranking effectiveness of RepLlama improved from 0 to a competitive score. This finding suggests that it is possible to convert an LLM into an effective  embedding model with little training data. On the other hand, \textit{PromptReps-dense only} achieved the best MRR@10 score on MSMARCO dev. However, the hybrid training loss coupled with hybrid retrieval achieved the highest effectiveness across different training settings on TREC DL; the only exception being the full-data setting on MSMARCO dev. These results demonstrate that PromptReps could be seen as a simple approach to obtaining a better initialization of LLM-based embedding models, which is more cost-effective than methods requiring further pre-training~\cite{behnamghader2024llm2vec,li2023making}.

\begin{table}[]\centering
	\caption{Supervised fine-tuning results}\label{tab:supervise}
	\resizebox{\columnwidth}{!}{
	\begin{tabular}{l|cccc}\hline
		&zero-shot &1k data &full data (490k) \\\hline
		&\multicolumn{3}{c}{MSMARCO dev MRR@10} \\\hline
		RepLlama3 &0.0 &27.88 &\textbf{42.77} \\
		PromptReps-dense only &17.54 &\textbf{28.48} &42.58 \\
		PromptReps-dense &17.54 &25.45 &41.86 \\
		PromptReps-sparse &20.85 &21.55 &34.15 \\
		PromptReps-hybrid &\textbf{24.62} &28.18 &42.48 \\\hline
		&\multicolumn{3}{c}{DL2019 NDCG@10} \\\hline
		RepLlama3 &0.0 &63.91 &73.19 \\
		PromptReps-dense only &50.62 &62.63 &73.50 \\
		PromptReps-dense &50.62 &64.48 &74.10 \\
		PromptReps-sparse &42.25 &47.23 &60.39 \\
		PromptReps-hybrid &\textbf{55.58} &\textbf{65.23} &\textbf{74.49} \\\hline
		&\multicolumn{3}{c}{DL2020 NDCG@10} \\\hline
		RepLlama3 &0.0 &63.10 &73.35 \\
		PromptReps-dense only &43.81 &61.46 &73.00 \\
		PromptReps-dense &43.81 &61.04 &73.65 \\
		PromptReps-sparse &45.60 &50.15 &62.81 \\
		PromptReps-hybrid &\textbf{56.66} &\textbf{64.01} &\textbf{73.87} \\
		\hline
	\end{tabular}
}
\end{table}

\section{Alternative Representation and Scoring}\label{sec:reps}
In the previous sections, we only considered using the representations (dense and sparse) yielded from the last token in the prompt for document retrieval. These representations, in the context of generative LLMs, are responsible for predicting the first generated token. We define this setting as \textit{First-token single-representation} (FTSR). We have demonstrated that this simple way of generating representations is effective for document retrieval; however, these representations might be sub-optimal. For example, LLMs use sub-word tokenization algorithms such as SentencePiece~\cite{kudo-richardson-2018-sentencepiece}. This tokenization might split a word into sub-words, meaning that the first generated token might just be a sub-word. Using the representation of the whole word might be a better representation than the first token representation. Additionally, previous works in multi-vector dense retrieval such as ColBERT~\cite{colbert} demonstrated that using multiple representations could be beneficial for document retrieval. How can we use PromptReps to also generate single-word representations or multiple representations that can potentially enhance the retrieval effectiveness? In this section, we explore these alternative representations.

%

\begin{figure*}[h]
    \centering
    \vspace{-10pt}
    \begin{minipage}{.5\textwidth}
        \centering
        \includegraphics[width=0.9\columnwidth]{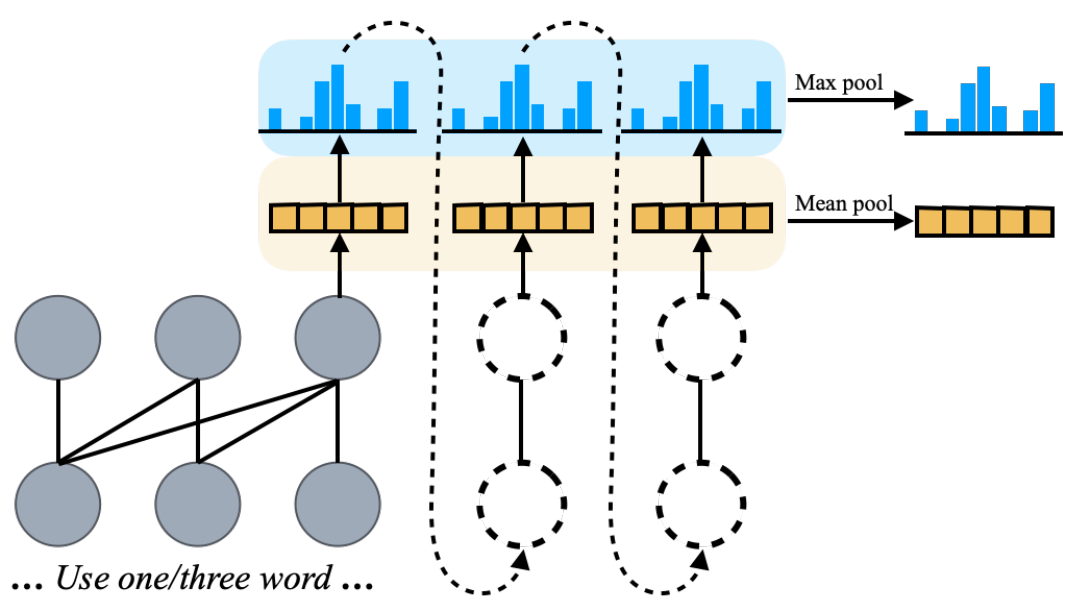}
	\caption{\textit{First-word single-representations} or \textit{Multi-token single-representation}.}
	\label{fig:single_rep}
    \end{minipage}%
    \begin{minipage}{0.5\textwidth}
        \centering
        \includegraphics[width=0.95\columnwidth]{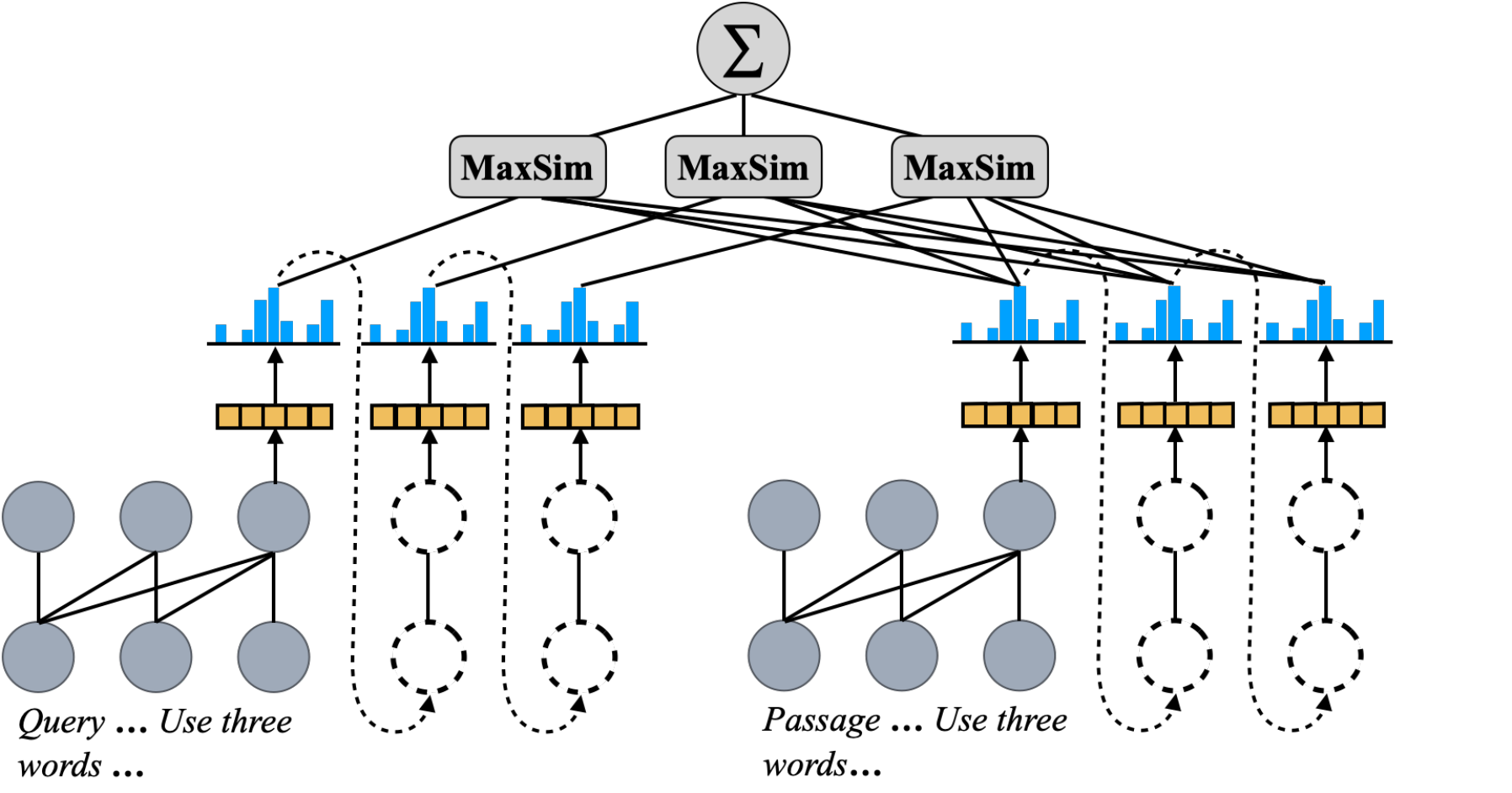}
	\caption{Multi-representations with ColBERT scoring.}
	\label{fig:colbert_rep}
    \end{minipage}
\end{figure*}

\begin{figure*}[h!]
	\vspace{-10pt}
	\centering
	\includegraphics[width=1\textwidth]{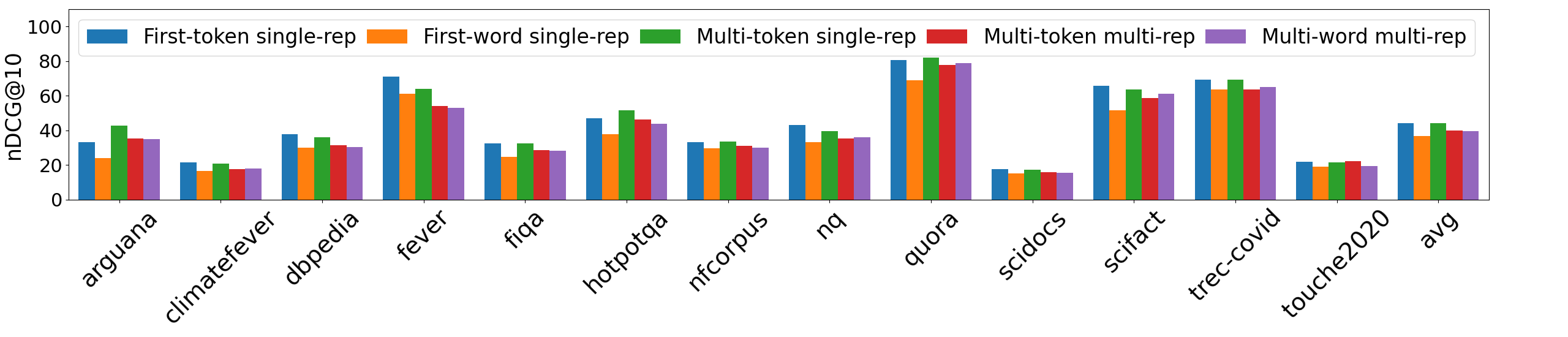}\vspace{-12pt}
	\caption{Hybrid retrieval results of different representation methods on BEIR.}
	\label{fig:rep_results}
	\vspace{-10pt}
\end{figure*}

\textit{First-word single-representation} (FWSR) and \textit{Multi-token single-representation} (MTSR). Instead of using the representations of the first generated token, these two methods let the LLM finish the generation\footnote{We simply use greedy generation.} of the whole word or multiple words, controlled by the given prompt (``\textit{Use one word}'' or ``\textit{Use three words}''), as illustrated in Figure~\ref{fig:single_rep}. The end of generation is detected by the token ` \textit{''} '. We then pool all the representations of the generated tokens to form a single dense and sparse representation for the input text. For the dense representation we use mean pooling and for the sparse representation we use max pooling. Once representations are obtained, the scoring is the same as FTSR.

\textit{Multi-token multi-representation} (MTMR) and \textit{Multi-word multi-representation} (MWMR). Instead of using a single representation for retrieval, these two methods prompt the LLM to generate multiple words and then index each generated representation separately. The difference between the two is that MTMR keeps all the token representations in the index, while MWMR first groups tokens into words by using space, and then creates a single representation for each word by using max pooling for sparse representations and mean pooling for dense representations. During retrieval, we follow the ColBERT scoring method where the relevance score of a document is computed by the sum of the maximum similarity of each query representation against each document representation (Figure~\ref{fig:colbert_rep}).

Hybrid retrieval results are shown in Figure~\ref{fig:rep_results}, and full dense and sparse retrieval results in Appendix~\ref{appendix:rep_results}. Results show that all the explored methods are able to perform document retrieval. The FTSR and MTSR generally perform the best. However, we note that MTSR requires more token generation steps and thus has higher query latency. The FWSR performs the worst, suggesting that sub-word representations hurt the retrieval performance for single-word generation prompts. On the other hand, multi-representation methods with ColBERT scoring methods do not seem beneficial. Thus, we conclude that the simplest FTSR is sufficient to represent the input text for document retrieval.

\section{Conclusion}
We introduced PromptReps, a simple yet effective method that prompts LLMs to generate dense and sparse representations for zero-shot document retrieval without any further training. We show that modern LLMs are effective text encoders by themselves, and prompt engineering is sufficient to stimulate their text encoding ability.

For future works, techniques like few-shot in-context learning~\cite{gpt3}, chain-of-thought prompting~\cite{cot}, and auto-prompt optimization methods~\cite{yang2024large,fernando2023promptbreeder}, which have proven to be effective in text-generation tasks, could potentially be leveraged here to enhance embedding generation.

Moreover, it has been shown that the instruction-following ability of LLMs could be transferred to embedding models with synthetic instruction fine-tuning data~\cite{wang2024improving}. In our work, we always keep the instruction prompt consistent across different IR tasks, which could be sub-optimal. It is interesting to investigate how to customize instructions for PromptReps to generate embeddings specific to different domains, tasks, or even to multi-lingual and cross-lingual IR settings.

Finally, our prompting method could be seen as a simple approach to obtaining a better initialization of LLM-based embedding models and all the previous contrastive pre-training with paired text data and synthetically generated data could be applied on top of our method and could potentially yield improved LLM-based embedding models.

\newpage
\section{Limitations}
PromptReps has higher query latency than other LLM-based dense retrievers if no further optimization is implemented.  This limitation comes from two aspects. 

First, although the computation of document representations happens offline thus will not affect query latency, the query representations are created online. PromptReps adds extra prompt texts on top of the query text thus has a longer input length -- and LLM inference time is proportional to prompt length. However, we believe this limitation can be mitigated by leveraging recent works on prompt compression to compress the fixed prompt tokens into few or even a single latent token~\cite{ge2024incontext,cheng2024xrag}.

Second, the highest effectiveness for PromptReps is achieved in the hybrid retrieval setting. Compared to previous works which use dense representations only, the hybrid setting requires both dense and sparse retrieval, thus the extra sparse retrieval introduces extra query latency (and requires additional disk/memory space for the inverted index). However, PromptReps actually only requires a limited query latency overhead if dense and sparse retrieval are implemented in parallel. In our method, obtaining both dense and sparse representations only requires a single LLM forward inference; the only extra computation is the dot product of the dense vector with the token embeddings, which is very fast on GPU. For document search, since we heavily sparsified the sparse representation, in our experiments, our sparse retriever is much faster than BM25, and the bottleneck is the dense retriever. Since the dense and sparse search could be run in parallel and the hybrid operation is a simple linear interpolation of both rankings (very fast on CPU), the query latency of the hybrid process only depends on the dense retrieval latency, and it is thus very close to previous methods.

\section{Ethical Considerations}
In our experiments, we use PromptReps coupled with LLMs with a large number of parameters (up to 70B in our experiments) to encode the BEIR and MSMARCO datasets, which contain millions of documents. Although no LLM training was conducted, we are aware that our experiments might still have consumed significant energy, thus contributing to CO2 emissions~\cite{scells2022reduce} and water consumption~\cite{zuccon2023beyond}.
In addition, since we leverage LLMs in a black-box manner and LLMs' generation might contain biases~\cite{gallegos2024bias}, the representations generated by LLMs may be biased towards certain contents or topics. Future work could consider how to mitigate biases in PromptReps via prompt engineering.

 \section*{Acknowledgments}
 The authors from the University of Waterloo acknowledge the support of the Natural Sciences and Engineering Research Council (NSERC) of Canada.

\bibliography{anthology,custom}

\begin{thebibliography}{66}
\providecommand{\natexlab}[1]{#1}

\bibitem[{Abdin et~al.(2024)}]{abdin2024phi3}
Marah Abdin et~al. 2024.
\newblock \href {https://arxiv.org/abs/2404.14219} {Phi-3 technical report: A
  highly capable language model locally on your phone}.
\newblock \emph{Preprint}, arXiv:2404.14219.

\bibitem[{AI@Meta(2024)}]{llama3modelcard}
AI@Meta. 2024.
\newblock \href {https://github.com/meta-llama/llama3/blob/main/MODEL_CARD.md}
  {Llama 3 model card}.

\bibitem[{Bajaj et~al.(2018)Bajaj, Campos, Craswell, Deng, Gao, Liu, Majumder,
  McNamara, Mitra, Nguyen, Rosenberg, Song, Stoica, Tiwary, and
  Wang}]{bajaj2018ms}
Payal Bajaj, Daniel Campos, Nick Craswell, Li~Deng, Jianfeng Gao, Xiaodong Liu,
  Rangan Majumder, Andrew McNamara, Bhaskar Mitra, Tri Nguyen, Mir Rosenberg,
  Xia Song, Alina Stoica, Saurabh Tiwary, and Tong Wang. 2018.
\newblock \href {https://arxiv.org/abs/1611.09268} {{MS MARCO}: A human
  generated machine reading comprehension dataset}.
\newblock \emph{Preprint}, arXiv:1611.09268.

\bibitem[{Bassani and Romelli(2022)}]{ranx}
Elias Bassani and Luca Romelli. 2022.
\newblock \href {https://doi.org/10.1145/3511808.3557207} {ranx.fuse: A
  {Python} library for metasearch}.
\newblock In \emph{Proceedings of the 31st ACM International Conference on
  Information \& Knowledge Management}, CIKM '22, pages 4808--4812, New York,
  NY, USA. Association for Computing Machinery.

\bibitem[{BehnamGhader et~al.(2024)BehnamGhader, Adlakha, Mosbach, Bahdanau,
  Chapados, and Reddy}]{behnamghader2024llm2vec}
Parishad BehnamGhader, Vaibhav Adlakha, Marius Mosbach, Dzmitry Bahdanau,
  Nicolas Chapados, and Siva Reddy. 2024.
\newblock \href {https://arxiv.org/abs/2404.05961} {Llm2vec: Large language
  models are secretly powerful text encoders}.
\newblock \emph{Preprint}, arXiv:2404.05961.

\bibitem[{Bird and Loper(2004)}]{bird-loper-2004-nltk}
Steven Bird and Edward Loper. 2004.
\newblock \href {https://aclanthology.org/P04-3031} {{NLTK}: The natural
  language toolkit}.
\newblock In \emph{Proceedings of the {ACL} Interactive Poster and
  Demonstration Sessions}, pages 214--217, Barcelona, Spain. Association for
  Computational Linguistics.

\bibitem[{Brown et~al.(2020)Brown, Mann, Ryder, Subbiah, Kaplan, Dhariwal,
  Neelakantan, Shyam, Sastry, Askell, Agarwal, Herbert-Voss, Krueger, Henighan,
  Child, Ramesh, Ziegler, Wu, Winter, Hesse, Chen, Sigler, Litwin, Gray, Chess,
  Clark, Berner, McCandlish, Radford, Sutskever, and Amodei}]{gpt3}
Tom Brown, Benjamin Mann, Nick Ryder, Melanie Subbiah, Jared~D Kaplan, Prafulla
  Dhariwal, Arvind Neelakantan, Pranav Shyam, Girish Sastry, Amanda Askell,
  Sandhini Agarwal, Ariel Herbert-Voss, Gretchen Krueger, Tom Henighan, Rewon
  Child, Aditya Ramesh, Daniel Ziegler, Jeffrey Wu, Clemens Winter, Chris
  Hesse, Mark Chen, Eric Sigler, Mateusz Litwin, Scott Gray, Benjamin Chess,
  Jack Clark, Christopher Berner, Sam McCandlish, Alec Radford, Ilya Sutskever,
  and Dario Amodei. 2020.
\newblock \href
  {https://proceedings.neurips.cc/paper_files/paper/2020/file/1457c0d6bfcb4967418bfb8ac142f64a-Paper.pdf}
  {Language models are few-shot learners}.
\newblock In \emph{Advances in Neural Information Processing Systems},
  volume~33, pages 1877--1901. Curran Associates, Inc.

\bibitem[{Cheng et~al.(2024)Cheng, Wang, Zhang, Ge, Chen, Wei, Zhang, and
  Zhao}]{cheng2024xrag}
Xin Cheng, Xun Wang, Xingxing Zhang, Tao Ge, Si-Qing Chen, Furu Wei, Huishuai
  Zhang, and Dongyan Zhao. 2024.
\newblock \href {https://arxiv.org/abs/2405.13792} {xrag: Extreme context
  compression for retrieval-augmented generation with one token}.
\newblock \emph{Preprint}, arXiv:2405.13792.

\bibitem[{Craswell et~al.(2020)Craswell, Mitra, Yilmaz, Campos, and
  Voorhees}]{craswell2020overview}
Nick Craswell, Bhaskar Mitra, Emine Yilmaz, Daniel Campos, and Ellen~M.
  Voorhees. 2020.
\newblock \href {https://arxiv.org/abs/2003.07820} {Overview of the {TREC} 2019
  deep learning track}.
\newblock \emph{Preprint}, arXiv:2003.07820.

\bibitem[{Douze et~al.(2024)Douze, Guzhva, Deng, Johnson, Szilvasy, Mazar{\'e},
  Lomeli, Hosseini, and J{\'e}gou}]{douze2024faiss}
Matthijs Douze, Alexandr Guzhva, Chengqi Deng, Jeff Johnson, Gergely Szilvasy,
  Pierre-Emmanuel Mazar{\'e}, Maria Lomeli, Lucas Hosseini, and Herv{\'e}
  J{\'e}gou. 2024.
\newblock \href {https://arxiv.org/abs/2401.08281} {The faiss library}.
\newblock \emph{Preprint}, arXiv:2401.08281.

\bibitem[{Fang et~al.(2024)Fang, Zhan, Ai, Mao, Su, Chen, and Liu}]{scalingDR}
Yan Fang, Jingtao Zhan, Qingyao Ai, Jiaxin Mao, Weihang Su, Jia Chen, and Yiqun
  Liu. 2024.
\newblock \href {https://doi.org/10.1145/3626772.3657743} {Scaling laws for
  dense retrieval}.
\newblock In \emph{Proceedings of the 47th International ACM SIGIR Conference
  on Research and Development in Information Retrieval}, SIGIR '24, page
  1339–1349, New York, NY, USA. Association for Computing Machinery.

\bibitem[{Fernando et~al.(2023)Fernando, Banarse, Michalewski, Osindero, and
  Rockt{\"a}schel}]{fernando2023promptbreeder}
Chrisantha Fernando, Dylan Banarse, Henryk Michalewski, Simon Osindero, and Tim
  Rockt{\"a}schel. 2023.
\newblock \href {https://arxiv.org/abs/2309.16797} {Promptbreeder:
  Self-referential self-improvement via prompt evolution}.
\newblock \emph{Preprint}, arXiv:2309.16797.

\bibitem[{Formal et~al.(2022)Formal, Lassance, Piwowarski, and
  Clinchant}]{splade++}
Thibault Formal, Carlos Lassance, Benjamin Piwowarski, and St\'{e}phane
  Clinchant. 2022.
\newblock \href {https://doi.org/10.1145/3477495.3531857} {From distillation to
  hard negative sampling: Making sparse neural {IR} models more effective}.
\newblock In \emph{Proceedings of the 45th International ACM SIGIR Conference
  on Research and Development in Information Retrieval}, SIGIR '22, page
  2353–2359, New York, NY, USA. Association for Computing Machinery.

\bibitem[{Formal et~al.(2021)Formal, Piwowarski, and Clinchant}]{splade}
Thibault Formal, Benjamin Piwowarski, and St\'{e}phane Clinchant. 2021.
\newblock \href {https://doi.org/10.1145/3404835.3463098} {{SPLADE}: Sparse
  lexical and expansion model for first stage ranking}.
\newblock In \emph{Proceedings of the 44th International ACM SIGIR Conference
  on Research and Development in Information Retrieval}, SIGIR '21, pages
  2288--2292, New York, NY, USA. Association for Computing Machinery.

\bibitem[{Gallegos et~al.(2024)Gallegos, Rossi, Barrow, Tanjim, Kim,
  Dernoncourt, Yu, Zhang, and Ahmed}]{gallegos2024bias}
Isabel~O. Gallegos, Ryan~A. Rossi, Joe Barrow, Md~Mehrab Tanjim, Sungchul Kim,
  Franck Dernoncourt, Tong Yu, Ruiyi Zhang, and Nesreen~K. Ahmed. 2024.
\newblock \href {https://doi.org/10.1162/coli_a_00524} {Bias and fairness in
  large language models: A survey}.
\newblock \emph{Computational Linguistics}, 50(3):1097--1179.

\bibitem[{Gao et~al.(2023{\natexlab{a}})Gao, Ma, Lin, and
  Callan}]{gao-etal-2023-precise}
Luyu Gao, Xueguang Ma, Jimmy Lin, and Jamie Callan. 2023{\natexlab{a}}.
\newblock \href {https://doi.org/10.18653/v1/2023.acl-long.99} {Precise
  zero-shot dense retrieval without relevance labels}.
\newblock In \emph{Proceedings of the 61st Annual Meeting of the Association
  for Computational Linguistics (Volume 1: Long Papers)}, pages 1762--1777,
  Toronto, Canada. Association for Computational Linguistics.

\bibitem[{Gao et~al.(2023{\natexlab{b}})Gao, Ma, Lin, and Callan}]{tevatron}
Luyu Gao, Xueguang Ma, Jimmy Lin, and Jamie Callan. 2023{\natexlab{b}}.
\newblock \href {https://doi.org/10.1145/3539618.3591805} {Tevatron: An
  efficient and flexible toolkit for neural retrieval}.
\newblock In \emph{Proceedings of the 46th International ACM SIGIR Conference
  on Research and Development in Information Retrieval}, SIGIR '23, pages
  3120--3124, New York, NY, USA. Association for Computing Machinery.

\bibitem[{Gao et~al.(2021)Gao, Yao, and Chen}]{gao-etal-2021-simcse}
Tianyu Gao, Xingcheng Yao, and Danqi Chen. 2021.
\newblock \href {https://doi.org/10.18653/v1/2021.emnlp-main.552} {{S}im{CSE}:
  Simple contrastive learning of sentence embeddings}.
\newblock In \emph{Proceedings of the 2021 Conference on Empirical Methods in
  Natural Language Processing}, pages 6894--6910, Online and Punta Cana,
  Dominican Republic. Association for Computational Linguistics.

\bibitem[{Ge et~al.(2024)Ge, Jing, Wang, Wang, Chen, and Wei}]{ge2024incontext}
Tao Ge, Hu~Jing, Lei Wang, Xun Wang, Si-Qing Chen, and Furu Wei. 2024.
\newblock \href {https://openreview.net/forum?id=uREj4ZuGJE} {In-context
  autoencoder for context compression in a large language model}.
\newblock In \emph{The Twelfth International Conference on Learning
  Representations}.

\bibitem[{Hendrycks et~al.(2021)Hendrycks, Burns, Basart, Zou, Mazeika, Song,
  and Steinhardt}]{hendryckstest2021}
Dan Hendrycks, Collin Burns, Steven Basart, Andy Zou, Mantas Mazeika, Dawn
  Song, and Jacob Steinhardt. 2021.
\newblock \href {https://openreview.net/forum?id=d7KBjmI3GmQ} {Measuring
  massive multitask language understanding}.
\newblock In \emph{International Conference on Learning Representations}.

\bibitem[{Hu et~al.(2022)Hu, yelong shen, Wallis, Allen-Zhu, Li, Wang, Wang,
  and Chen}]{hu2022lora}
Edward~J Hu, yelong shen, Phillip Wallis, Zeyuan Allen-Zhu, Yuanzhi Li, Shean
  Wang, Lu~Wang, and Weizhu Chen. 2022.
\newblock \href {https://openreview.net/forum?id=nZeVKeeFYf9} {Lo{RA}: Low-rank
  adaptation of large language models}.
\newblock In \emph{International Conference on Learning Representations}.

\bibitem[{Izacard et~al.(2022)Izacard, Caron, Hosseini, Riedel, Bojanowski,
  Joulin, and Grave}]{izacard2022unsupervised}
Gautier Izacard, Mathilde Caron, Lucas Hosseini, Sebastian Riedel, Piotr
  Bojanowski, Armand Joulin, and Edouard Grave. 2022.
\newblock \href {https://arxiv.org/abs/2112.09118} {Unsupervised dense
  information retrieval with contrastive learning}.
\newblock \emph{Preprint}, arXiv:2112.09118.

\bibitem[{Jiang et~al.(2023{\natexlab{a}})Jiang, Sablayrolles, Mensch, Bamford,
  Chaplot, de~las Casas, Bressand, Lengyel, Lample, Saulnier, Lavaud, Lachaux,
  Stock, Scao, Lavril, Wang, Lacroix, and Sayed}]{jiang2023mistral}
Albert~Q. Jiang, Alexandre Sablayrolles, Arthur Mensch, Chris Bamford,
  Devendra~Singh Chaplot, Diego de~las Casas, Florian Bressand, Gianna Lengyel,
  Guillaume Lample, Lucile Saulnier, L{\'e}lio~Renard Lavaud, Marie-Anne
  Lachaux, Pierre Stock, Teven~Le Scao, Thibaut Lavril, Thomas Wang,
  Timoth{\'e}e Lacroix, and William~El Sayed. 2023{\natexlab{a}}.
\newblock \href {https://arxiv.org/abs/2310.06825} {Mistral 7b}.
\newblock \emph{Preprint}, arXiv:2310.06825.

\bibitem[{Jiang et~al.(2023{\natexlab{b}})Jiang, Huang, Luan, Wang, and
  Zhuang}]{jiang2023scaling}
Ting Jiang, Shaohan Huang, Zhongzhi Luan, Deqing Wang, and Fuzhen Zhuang.
  2023{\natexlab{b}}.
\newblock \href {https://arxiv.org/abs/2307.16645} {Scaling sentence embeddings
  with large language models}.
\newblock \emph{Preprint}, arXiv:2307.16645.

\bibitem[{Kaplan et~al.(2020)Kaplan, McCandlish, Henighan, Brown, Chess, Child,
  Gray, Radford, Wu, and Amodei}]{kaplan2020scaling}
Jared Kaplan, Sam McCandlish, Tom Henighan, Tom~B. Brown, Benjamin Chess, Rewon
  Child, Scott Gray, Alec Radford, Jeffrey Wu, and Dario Amodei. 2020.
\newblock \href {https://arxiv.org/abs/2001.08361} {Scaling laws for neural
  language models}.
\newblock \emph{Preprint}, arXiv:2001.08361.

\bibitem[{Karpukhin et~al.(2020)Karpukhin, Oguz, Min, Lewis, Wu, Edunov, Chen,
  and Yih}]{dpr}
Vladimir Karpukhin, Barlas Oguz, Sewon Min, Patrick Lewis, Ledell Wu, Sergey
  Edunov, Danqi Chen, and Wen-tau Yih. 2020.
\newblock \href {https://doi.org/10.18653/v1/2020.emnlp-main.550} {Dense
  passage retrieval for open-domain question answering}.
\newblock In \emph{Proceedings of the 2020 Conference on Empirical Methods in
  Natural Language Processing (EMNLP)}, pages 6769--6781, Online. Association
  for Computational Linguistics.

\bibitem[{Khattab and Zaharia(2020)}]{colbert}
Omar Khattab and Matei Zaharia. 2020.
\newblock \href {https://doi.org/10.1145/3397271.3401075} {{ColBERT}: Efficient
  and effective passage search via contextualized late interaction over bert}.
\newblock In \emph{Proceedings of the 43rd International ACM SIGIR Conference
  on Research and Development in Information Retrieval}, SIGIR '20, pages
  39--48, New York, NY, USA. Association for Computing Machinery.

\bibitem[{Khramtsova et~al.(2023)Khramtsova, Zhuang, Baktashmotlagh, Wang, and
  Zuccon}]{khramtsova2023selecting}
Ekaterina Khramtsova, Shengyao Zhuang, Mahsa Baktashmotlagh, Xi~Wang, and Guido
  Zuccon. 2023.
\newblock \href {https://doi.org/10.1145/3624918.3625330} {Selecting which
  dense retriever to use for zero-shot search}.
\newblock In \emph{Proceedings of the Annual International ACM SIGIR Conference
  on Research and Development in Information Retrieval in the Asia Pacific
  Region}, SIGIR-AP '23, page 223–233, New York, NY, USA. Association for
  Computing Machinery.

\bibitem[{Khramtsova et~al.(2024)Khramtsova, Zhuang, Baktashmotlagh, and
  Zuccon}]{khramtsova2024leveraging}
Ekaterina Khramtsova, Shengyao Zhuang, Mahsa Baktashmotlagh, and Guido Zuccon.
  2024.
\newblock \href {https://doi.org/10.1145/3626772.3657798} {Leveraging {LLMs}
  for unsupervised dense retriever ranking}.
\newblock In \emph{Proceedings of the 47th International ACM SIGIR Conference
  on Research and Development in Information Retrieval}, SIGIR '24, page
  1307–1317, New York, NY, USA. Association for Computing Machinery.

\bibitem[{Kudo and Richardson(2018)}]{kudo-richardson-2018-sentencepiece}
Taku Kudo and John Richardson. 2018.
\newblock \href {https://doi.org/10.18653/v1/D18-2012} {{S}entence{P}iece: A
  simple and language independent subword tokenizer and detokenizer for neural
  text processing}.
\newblock In \emph{Proceedings of the 2018 Conference on Empirical Methods in
  Natural Language Processing: System Demonstrations}, pages 66--71, Brussels,
  Belgium. Association for Computational Linguistics.

\bibitem[{Lee et~al.(2024)Lee, Dai, Ren, Chen, Cer, Cole, Hui, Boratko,
  Kapadia, Ding, Luan, Duddu, Abrego, Shi, Gupta, Kusupati, Jain, Jonnalagadda,
  Chang, and Naim}]{lee2024gecko}
Jinhyuk Lee, Zhuyun Dai, Xiaoqi Ren, Blair Chen, Daniel Cer, Jeremy~R. Cole,
  Kai Hui, Michael Boratko, Rajvi Kapadia, Wen Ding, Yi~Luan, Sai Meher~Karthik
  Duddu, Gustavo~Hernandez Abrego, Weiqiang Shi, Nithi Gupta, Aditya Kusupati,
  Prateek Jain, Siddhartha~Reddy Jonnalagadda, Ming-Wei Chang, and Iftekhar
  Naim. 2024.
\newblock \href {https://arxiv.org/abs/2403.20327} {Gecko: Versatile text
  embeddings distilled from large language models}.
\newblock \emph{Preprint}, arXiv:2403.20327.

\bibitem[{Lei et~al.(2024)Lei, Wu, Zhou, Shen, Cao, Tao, and
  Yates}]{lei2024metatask}
Yibin Lei, Di~Wu, Tianyi Zhou, Tao Shen, Yu~Cao, Chongyang Tao, and Andrew
  Yates. 2024.
\newblock \href {https://arxiv.org/abs/2402.18458} {Meta-task prompting elicits
  embedding from large language models}.
\newblock \emph{Preprint}, arXiv:2402.18458.

\bibitem[{Li et~al.(2023)Li, Liu, Xiao, and Shao}]{li2023making}
Chaofan Li, Zheng Liu, Shitao Xiao, and Yingxia Shao. 2023.
\newblock \href {https://arxiv.org/abs/2312.15503} {Making large language
  models a better foundation for dense retrieval}.
\newblock \emph{Preprint}, arXiv:2312.15503.

\bibitem[{Liao et~al.(2024)Liao, Li, Yang, Wu, Yuan, Wang, and
  He}]{liao2024llara}
Jiayi Liao, Sihang Li, Zhengyi Yang, Jiancan Wu, Yancheng Yuan, Xiang Wang, and
  Xiangnan He. 2024.
\newblock \href {https://arxiv.org/abs/2312.02445} {Llara: Large
  language-recommendation assistant}.
\newblock \emph{Preprint}, arXiv:2312.02445.

\bibitem[{Lin and Ma(2021)}]{lin2021brief}
Jimmy Lin and Xueguang Ma. 2021.
\newblock \href {https://arxiv.org/abs/2106.14807} {A few brief notes on
  {DeepImpact}, {COIL}, and a conceptual framework for information retrieval
  techniques}.
\newblock \emph{Preprint}, arXiv:2106.14807.

\bibitem[{Lin et~al.(2021)Lin, Ma, Lin, Yang, Pradeep, and Nogueira}]{pyserini}
Jimmy Lin, Xueguang Ma, Sheng-Chieh Lin, Jheng-Hong Yang, Ronak Pradeep, and
  Rodrigo Nogueira. 2021.
\newblock \href {https://doi.org/10.1145/3404835.3463238} {Pyserini: A {Python}
  toolkit for reproducible information retrieval research with sparse and dense
  representations}.
\newblock In \emph{Proceedings of the 44th International ACM SIGIR Conference
  on Research and Development in Information Retrieval}, SIGIR '21, pages
  2356--2362, New York, NY, USA. Association for Computing Machinery.

\bibitem[{Lin et~al.(2023)Lin, Asai, Li, Oguz, Lin, Mehdad, Yih, and
  Chen}]{lin2023train}
Sheng-Chieh Lin, Akari Asai, Minghan Li, Barlas Oguz, Jimmy Lin, Yashar Mehdad,
  Wen-tau Yih, and Xilun Chen. 2023.
\newblock \href {https://doi.org/10.18653/v1/2023.findings-emnlp.423} {How to
  train your {Dragon}: Diverse augmentation towards generalizable dense
  retrieval}.
\newblock In \emph{Findings of the Association for Computational Linguistics:
  EMNLP 2023}, pages 6385--6400, Singapore. Association for Computational
  Linguistics.

\bibitem[{Liu et~al.(2023)Liu, Han, Ma, Zhang, Yang, Tian, He, Li, He, Liu, Wu,
  Zhao, Zhu, Li, Qiang, Shen, Liu, and Ge}]{Liu_2023}
Yiheng Liu, Tianle Han, Siyuan Ma, Jiayue Zhang, Yuanyuan Yang, Jiaming Tian,
  Hao He, Antong Li, Mengshen He, Zhengliang Liu, Zihao Wu, Lin Zhao, Dajiang
  Zhu, Xiang Li, Ning Qiang, Dingang Shen, Tianming Liu, and Bao Ge. 2023.
\newblock \href {https://doi.org/10.1016/j.metrad.2023.100017} {Summary of
  {ChatGPT}-related research and perspective towards the future of large
  language models}.
\newblock \emph{Meta-Radiology}, 1(2):100017.

\bibitem[{Lupart et~al.(2022)Lupart, Formal, and
  Clinchant}]{Lupart2022MSShiftAA}
Simon Lupart, Thibault Formal, and St{\'e}phane Clinchant. 2022.
\newblock \href {https://api.semanticscholar.org/CorpusID:256231516}
  {{MS-Shift}: An analysis of {MS MARCO} distribution shifts on neural
  retrieval}.
\newblock In \emph{European Conference on Information Retrieval}.

\bibitem[{Ma et~al.(2024)Ma, Wang, Yang, Wei, and Lin}]{ma2023finetuning}
Xueguang Ma, Liang Wang, Nan Yang, Furu Wei, and Jimmy Lin. 2024.
\newblock \href {https://doi.org/10.1145/3626772.3657951} {Fine-tuning {LLaMA}
  for multi-stage text retrieval}.
\newblock In \emph{Proceedings of the 47th International ACM SIGIR Conference
  on Research and Development in Information Retrieval}, SIGIR '24, page
  2421–2425, New York, NY, USA. Association for Computing Machinery.

\bibitem[{Ma et~al.(2023)Ma, Zhang, Pradeep, and Lin}]{ma2023zeroshot}
Xueguang Ma, Xinyu Zhang, Ronak Pradeep, and Jimmy Lin. 2023.
\newblock \href {https://arxiv.org/abs/2305.02156} {Zero-shot listwise document
  reranking with a large language model}.
\newblock \emph{Preprint}, arXiv:2305.02156.

\bibitem[{Mallia et~al.(2021)Mallia, Khattab, Suel, and
  Tonellotto}]{deepimpact}
Antonio Mallia, Omar Khattab, Torsten Suel, and Nicola Tonellotto. 2021.
\newblock \href {https://doi.org/10.1145/3404835.3463030} {Learning passage
  impacts for inverted indexes}.
\newblock In \emph{Proceedings of the 44th International ACM SIGIR Conference
  on Research and Development in Information Retrieval}, SIGIR '21, pages
  1723--1727, New York, NY, USA. Association for Computing Machinery.

\bibitem[{Open\-AI(2024)}]{openai2023gpt4}
Open\-AI. 2024.
\newblock \href {https://arxiv.org/abs/2303.08774} {{GPT-4} technical report}.
\newblock \emph{Preprint}, arXiv:2303.08774.

\bibitem[{Pradeep et~al.(2023)Pradeep, Sharifymoghaddam, and
  Lin}]{pradeep2023rankvicuna}
Ronak Pradeep, Sahel Sharifymoghaddam, and Jimmy Lin. 2023.
\newblock \href {https://arxiv.org/abs/2309.15088} {{RankVicuna}: Zero-shot
  listwise document reranking with open-source large language models}.
\newblock \emph{Preprint}, arXiv:2309.15088.

\bibitem[{Qin et~al.(2024)Qin, Jagerman, Hui, Zhuang, Wu, Yan, Shen, Liu, Liu,
  Metzler, Wang, and Bendersky}]{qin2024large}
Zhen Qin, Rolf Jagerman, Kai Hui, Honglei Zhuang, Junru Wu, Le~Yan, Jiaming
  Shen, Tianqi Liu, Jialu Liu, Donald Metzler, Xuanhui Wang, and Michael
  Bendersky. 2024.
\newblock \href {https://arxiv.org/abs/2306.17563} {Large language models are
  effective text rankers with pairwise ranking prompting}.
\newblock \emph{Preprint}, arXiv:2306.17563.

\bibitem[{Ren et~al.(2023)Ren, Qu, Liu, Zhao, Wu, Ding, Wu, Wang, and
  Wen}]{ren-etal-2023-thorough}
Ruiyang Ren, Yingqi Qu, Jing Liu, Xin Zhao, Qifei Wu, Yuchen Ding, Hua Wu,
  Haifeng Wang, and Ji-Rong Wen. 2023.
\newblock \href {https://doi.org/10.18653/v1/2023.findings-emnlp.1057} {A
  thorough examination on zero-shot dense retrieval}.
\newblock In \emph{Findings of the Association for Computational Linguistics:
  EMNLP 2023}, pages 15783--15796, Singapore. Association for Computational
  Linguistics.

\bibitem[{Sachan et~al.(2022)Sachan, Lewis, Joshi, Aghajanyan, Yih, Pineau, and
  Zettlemoyer}]{sachan-etal-2022-improving}
Devendra Sachan, Mike Lewis, Mandar Joshi, Armen Aghajanyan, Wen-tau Yih,
  Joelle Pineau, and Luke Zettlemoyer. 2022.
\newblock \href {https://doi.org/10.18653/v1/2022.emnlp-main.249} {Improving
  passage retrieval with zero-shot question generation}.
\newblock In \emph{Proceedings of the 2022 Conference on Empirical Methods in
  Natural Language Processing}, pages 3781--3797, Abu Dhabi, United Arab
  Emirates. Association for Computational Linguistics.

\bibitem[{Scells et~al.(2022)Scells, Zhuang, and Zuccon}]{scells2022reduce}
Harrisen Scells, Shengyao Zhuang, and Guido Zuccon. 2022.
\newblock \href {https://doi.org/10.1145/3477495.3531766} {Reduce, reuse,
  recycle: Green information retrieval research}.
\newblock In \emph{Proceedings of the 45th International ACM SIGIR Conference
  on Research and Development in Information Retrieval}, SIGIR '22, page
  2825–2837, New York, NY, USA. Association for Computing Machinery.

\bibitem[{Sun et~al.(2023)Sun, Yan, Ma, Wang, Ren, Chen, Yin, and
  Ren}]{Sun2023IsCG}
Weiwei Sun, Lingyong Yan, Xinyu Ma, Shuaiqiang Wang, Pengjie Ren, Zhumin Chen,
  Dawei Yin, and Zhaochun Ren. 2023.
\newblock \href {https://doi.org/10.18653/v1/2023.emnlp-main.923} {Is
  {C}hat{GPT} good at search? investigating large language models as re-ranking
  agents}.
\newblock In \emph{Proceedings of the 2023 Conference on Empirical Methods in
  Natural Language Processing}, pages 14918--14937, Singapore. Association for
  Computational Linguistics.

\bibitem[{Thakur et~al.(2021)Thakur, Reimers, R{\"u}ckl{\'e}, Srivastava, and
  Gurevych}]{thakur2021beir}
Nandan Thakur, Nils Reimers, Andreas R{\"u}ckl{\'e}, Abhishek Srivastava, and
  Iryna Gurevych. 2021.
\newblock \href {https://openreview.net/forum?id=wCu6T5xFjeJ} {{BEIR}: A
  heterogeneous benchmark for zero-shot evaluation of information retrieval
  models}.
\newblock In \emph{Thirty-fifth Conference on Neural Information Processing
  Systems Datasets and Benchmarks Track (Round 2)}.

\bibitem[{Touvron et~al.(2023)Touvron, Martin, Stone, Albert, Almahairi,
  Babaei, Bashlykov, Batra, Bhargava, Bhosale, Bikel, Blecher, Ferrer, Chen,
  Cucurull, Esiobu, Fernandes, Fu, Fu, Fuller, Gao, Goswami, Goyal, Hartshorn,
  Hosseini, Hou, Inan, Kardas, Kerkez, Khabsa, Kloumann, Korenev, Koura,
  Lachaux, Lavril, Lee, Liskovich, Lu, Mao, Martinet, Mihaylov, Mishra,
  Molybog, Nie, Poulton, Reizenstein, Rungta, Saladi, Schelten, Silva, Smith,
  Subramanian, Tan, Tang, Taylor, Williams, Kuan, Xu, Yan, Zarov, Zhang, Fan,
  Kambadur, Narang, Rodriguez, Stojnic, Edunov, and Scialom}]{llama2}
Hugo Touvron, Louis Martin, Kevin Stone, Peter Albert, Amjad Almahairi, Yasmine
  Babaei, Nikolay Bashlykov, Soumya Batra, Prajjwal Bhargava, Shruti Bhosale,
  Dan Bikel, Lukas Blecher, Cristian~Canton Ferrer, Moya Chen, Guillem
  Cucurull, David Esiobu, Jude Fernandes, Jeremy Fu, Wenyin Fu, Brian Fuller,
  Cynthia Gao, Vedanuj Goswami, Naman Goyal, Anthony Hartshorn, Saghar
  Hosseini, Rui Hou, Hakan Inan, Marcin Kardas, Viktor Kerkez, Madian Khabsa,
  Isabel Kloumann, Artem Korenev, Punit~Singh Koura, Marie-Anne Lachaux,
  Thibaut Lavril, Jenya Lee, Diana Liskovich, Yinghai Lu, Yuning Mao, Xavier
  Martinet, Todor Mihaylov, Pushkar Mishra, Igor Molybog, Yixin Nie, Andrew
  Poulton, Jeremy Reizenstein, Rashi Rungta, Kalyan Saladi, Alan Schelten, Ruan
  Silva, Eric~Michael Smith, Ranjan Subramanian, Xiaoqing~Ellen Tan, Binh Tang,
  Ross Taylor, Adina Williams, Jian~Xiang Kuan, Puxin Xu, Zheng Yan, Iliyan
  Zarov, Yuchen Zhang, Angela Fan, Melanie Kambadur, Sharan Narang, Aurelien
  Rodriguez, Robert Stojnic, Sergey Edunov, and Thomas Scialom. 2023.
\newblock \href {https://arxiv.org/abs/2307.09288} {Llama 2: Open foundation
  and fine-tuned chat models}.
\newblock \emph{Preprint}, arXiv:2307.09288.

\bibitem[{Wang et~al.(2024{\natexlab{a}})Wang, Yang, Huang, Jiao, Yang, Jiang,
  Majumder, and Wei}]{wang2024text}
Liang Wang, Nan Yang, Xiaolong Huang, Binxing Jiao, Linjun Yang, Daxin Jiang,
  Rangan Majumder, and Furu Wei. 2024{\natexlab{a}}.
\newblock \href {https://arxiv.org/abs/2212.03533} {Text embeddings by
  weakly-supervised contrastive pre-training}.
\newblock \emph{Preprint}, arXiv:2212.03533.

\bibitem[{Wang et~al.(2024{\natexlab{b}})Wang, Yang, Huang, Yang, Majumder, and
  Wei}]{wang2024improving}
Liang Wang, Nan Yang, Xiaolong Huang, Linjun Yang, Rangan Majumder, and Furu
  Wei. 2024{\natexlab{b}}.
\newblock \href {https://arxiv.org/abs/2401.00368} {Improving text embeddings
  with large language models}.
\newblock \emph{Preprint}, arXiv:2401.00368.

\bibitem[{Wang et~al.(2021)Wang, Zhuang, and Zuccon}]{bm25interpolation}
Shuai Wang, Shengyao Zhuang, and Guido Zuccon. 2021.
\newblock \href {https://doi.org/10.1145/3471158.3472233} {{BERT}-based dense
  retrievers require interpolation with bm25 for effective passage retrieval}.
\newblock In \emph{Proceedings of the 2021 ACM SIGIR International Conference
  on Theory of Information Retrieval}, ICTIR '21, pages 317--324, New York, NY,
  USA. Association for Computing Machinery.

\bibitem[{Wei et~al.(2022)Wei, Wang, Schuurmans, Bosma, ichter, Xia, Chi, Le,
  and Zhou}]{cot}
Jason Wei, Xuezhi Wang, Dale Schuurmans, Maarten Bosma, brian ichter, Fei Xia,
  Ed~Chi, Quoc~V Le, and Denny Zhou. 2022.
\newblock \href
  {https://proceedings.neurips.cc/paper_files/paper/2022/file/9d5609613524ecf4f15af0f7b31abca4-Paper-Conference.pdf}
  {Chain-of-thought prompting elicits reasoning in large language models}.
\newblock In \emph{Advances in Neural Information Processing Systems},
  volume~35, pages 24824--24837. Curran Associates, Inc.

\bibitem[{Wolf et~al.(2020)Wolf, Debut, Sanh, Chaumond, Delangue, Moi, Cistac,
  Rault, Louf, Funtowicz, Davison, Shleifer, von Platen, Ma, Jernite, Plu, Xu,
  Le~Scao, Gugger, Drame, Lhoest, and Rush}]{wolf-etal-2020-transformers}
Thomas Wolf, Lysandre Debut, Victor Sanh, Julien Chaumond, Clement Delangue,
  Anthony Moi, Pierric Cistac, Tim Rault, Remi Louf, Morgan Funtowicz, Joe
  Davison, Sam Shleifer, Patrick von Platen, Clara Ma, Yacine Jernite, Julien
  Plu, Canwen Xu, Teven Le~Scao, Sylvain Gugger, Mariama Drame, Quentin Lhoest,
  and Alexander Rush. 2020.
\newblock \href {https://doi.org/10.18653/v1/2020.emnlp-demos.6} {Transformers:
  State-of-the-art natural language processing}.
\newblock In \emph{Proceedings of the 2020 Conference on Empirical Methods in
  Natural Language Processing: System Demonstrations}, pages 38--45, Online.
  Association for Computational Linguistics.

\bibitem[{Xiong et~al.(2021)Xiong, Xiong, Li, Tang, Liu, Bennett, Ahmed, and
  Overwijk}]{ance}
Lee Xiong, Chenyan Xiong, Ye~Li, Kwok-Fung Tang, Jialin Liu, Paul~N. Bennett,
  Junaid Ahmed, and Arnold Overwijk. 2021.
\newblock \href {https://openreview.net/forum?id=zeFrfgyZln} {Approximate
  nearest neighbor negative contrastive learning for dense text retrieval}.
\newblock In \emph{International Conference on Learning Representations}.

\bibitem[{Yang et~al.(2024)Yang, Wang, Lu, Liu, Le, Zhou, and
  Chen}]{yang2024large}
Chengrun Yang, Xuezhi Wang, Yifeng Lu, Hanxiao Liu, Quoc~V Le, Denny Zhou, and
  Xinyun Chen. 2024.
\newblock \href {https://openreview.net/forum?id=Bb4VGOWELI} {Large language
  models as optimizers}.
\newblock In \emph{The Twelfth International Conference on Learning
  Representations}.

\bibitem[{Zhang et~al.(2024)Zhang, Chang, and Li}]{zhang2024simple}
Bowen Zhang, Kehua Chang, and Chunping Li. 2024.
\newblock \href {https://arxiv.org/abs/2404.03921} {Simple techniques for
  enhancing sentence embeddings in generative language models}.
\newblock \emph{Preprint}, arXiv:2404.03921.

\bibitem[{Zhuang et~al.(2023)Zhuang, Liu, Koopman, and
  Zuccon}]{zhuang-etal-2023-open}
Shengyao Zhuang, Bing Liu, Bevan Koopman, and Guido Zuccon. 2023.
\newblock \href {https://doi.org/10.18653/v1/2023.findings-emnlp.590}
  {Open-source large language models are strong zero-shot query likelihood
  models for document ranking}.
\newblock In \emph{Findings of the Association for Computational Linguistics:
  EMNLP 2023}, pages 8807--8817, Singapore. Association for Computational
  Linguistics.

\bibitem[{Zhuang et~al.(2024)Zhuang, Zhuang, Koopman, and
  Zuccon}]{zhuang2023setwise}
Shengyao Zhuang, Honglei Zhuang, Bevan Koopman, and Guido Zuccon. 2024.
\newblock \href {https://doi.org/10.1145/3626772.3657813} {A setwise approach
  for effective and highly efficient zero-shot ranking with large language
  models}.
\newblock In \emph{Proceedings of the 47th International ACM SIGIR Conference
  on Research and Development in Information Retrieval}, SIGIR '24, page
  38–47, New York, NY, USA. Association for Computing Machinery.

\bibitem[{Zhuang and Zuccon(2021{\natexlab{a}})}]{zhuang2021dealing}
Shengyao Zhuang and Guido Zuccon. 2021{\natexlab{a}}.
\newblock \href {https://doi.org/10.18653/v1/2021.emnlp-main.225} {Dealing with
  typos for {BERT}-based passage retrieval and ranking}.
\newblock In \emph{Proceedings of the 2021 Conference on Empirical Methods in
  Natural Language Processing}, pages 2836--2842, Online and Punta Cana,
  Dominican Republic. Association for Computational Linguistics.

\bibitem[{Zhuang and Zuccon(2021{\natexlab{b}})}]{zhuang2021fast}
Shengyao Zhuang and Guido Zuccon. 2021{\natexlab{b}}.
\newblock \href {https://arxiv.org/abs/2108.08513} {Fast passage re-ranking
  with contextualized exact term matching and efficient passage expansion}.
\newblock \emph{Preprint}, arXiv:2108.08513.

\bibitem[{Zhuang and Zuccon(2021{\natexlab{c}})}]{tilde}
Shengyao Zhuang and Guido Zuccon. 2021{\natexlab{c}}.
\newblock \href {https://doi.org/10.1145/3404835.3462922} {{TILDE}: Term
  independent likelihood model for passage re-ranking}.
\newblock In \emph{Proceedings of the 44th International ACM SIGIR Conference
  on Research and Development in Information Retrieval}, SIGIR '21, pages
  1483--1492, New York, NY, USA. Association for Computing Machinery.

\bibitem[{Zhuang and Zuccon(2022)}]{zhuang2022char}
Shengyao Zhuang and Guido Zuccon. 2022.
\newblock \href {https://doi.org/10.1145/3477495.3531951} {Characterbert and
  self-teaching for improving the robustness of dense retrievers on queries
  with typos}.
\newblock In \emph{Proceedings of the 45th International ACM SIGIR Conference
  on Research and Development in Information Retrieval}, SIGIR '22, page
  1444–1454, New York, NY, USA. Association for Computing Machinery.

\bibitem[{Zuccon et~al.(2023)Zuccon, Scells, and Zhuang}]{zuccon2023beyond}
Guido Zuccon, Harrisen Scells, and Shengyao Zhuang. 2023.
\newblock \href {https://doi.org/10.1145/3578337.3605121} {Beyond {CO2}
  emissions: The overlooked impact of water consumption of information
  retrieval models}.
\newblock In \emph{Proceedings of the 2023 ACM SIGIR International Conference
  on Theory of Information Retrieval}, ICTIR '23, page 283–289, New York, NY,
  USA. Association for Computing Machinery.

\end{thebibliography}

 \appendix

\section{Python code example}\label{appendix:python}
In Table~\ref{table:python} we provide a Python code implementation of PromptReps with Meta-Llama-3-8B-Instruct LLM. The example uses Huggingface transformers library (v4.40.1) with torch (v2.3.0), numpy (v1.26.4) and nltk (v3.9.1) to generate dense and sparse representations.
\begin{table*}[]
	\centering
	\caption{Python code example of generating dense and sparse representations with PromptReps.}\label{table:python}
\begin{adjustbox}{minipage=\textwidth, scale=1.0, margin=0cm 0cm 0cm 0cm}
\begin{lstlisting}
from transformers import AutoModelForCausalLM, AutoTokenizer
import torch
import numpy as np
from nltk import word_tokenize
from nltk.corpus import stopwords
import string
stopwords = set(stopwords.words('english') + list(string.punctuation))

model_id = 'meta-llama/Meta-Llama-3-8B-Instruct'
tokenizer = AutoTokenizer.from_pretrained(model_id)
model = AutoModelForCausalLM.from_pretrained(model_id).to('cuda') 

passage = 'The quick brown fox jumps over the lazy dog.'
messages = [
	{'role': 'system', 'content': 'You are an AI assistant that can understand human language.'},
	{'role': 'user', 'content': f'Passage: "{passage}". Use one word to represent the passage'
		                  	  f' in a retrieval task. Make sure your word is in lowercase.'},
	{'role': 'assistant', 'content': 'The word is "'}
]

input_ids = tokenizer.apply_chat_template(
			messages,
			add_generation_prompt=False,
			return_tensors='pt'
		)[:, :-1].to('cuda')  # the last special token is removed

outputs = model(input_ids=input_ids, return_dict=True, output_hidden_states=True)

# generating dense representation
dense_representation = outputs.hidden_states[-1][:, -1, :][0]

# generating sparse representation, log and relu the values
next_token_logits = torch.log(1 + torch.relu(outputs.logits))[:, -1, :][0]

# lower case and stopwords removal
words_in_text = [word for word in word_tokenize(passage.lower()) if word not in stopwords]

# extract token ids in the given passage
token_ids_in_text = set()
for word in words_in_text:
	token_ids_in_text.update(tokenizer.encode(word, add_special_tokens=False))
	
token_ids_in_text = torch.tensor(list(token_ids_in_text))

# get top tokens and quantization
top_k = min(len(token_ids_in_text), 128)
top_k_values, top_k_indices = next_token_logits[token_ids_in_text].topk(top_k, dim=-1)
values = np.rint(top_k_values.cpu().detach().float().numpy() * 100).astype(int)
tokens = [tokenizer.decode(i) for i in token_ids_in_text[top_k_indices.cpu().detach().float().numpy()]]

# final sparse representation
print({token: value for token, value in zip(tokens, values)})
# {'fox': 312, 'dog': 280, 'brown': 276, 'j': 273, 'quick': 265, 'lazy': 257, 'umps': 144}
\end{lstlisting}
\end{adjustbox}
\end{table*}

\section{Impact of Hybrid weights}\label{appendix:weights}
\begin{figure*}
	\centering
	\includegraphics[width=0.7\linewidth]{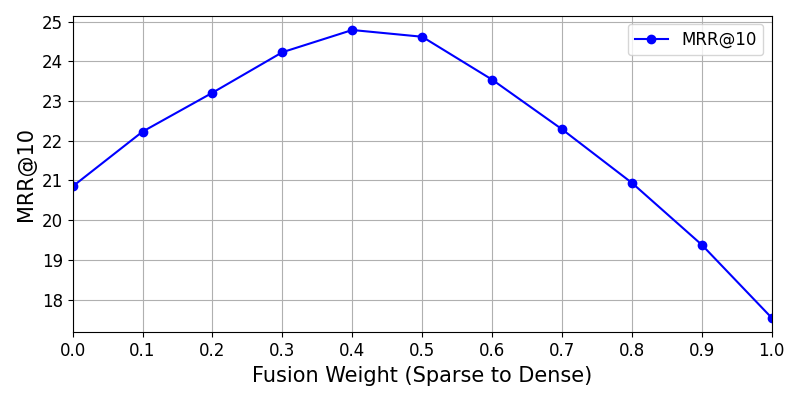}
	\caption{Average MRR@10 scores on MSMARCO dev queries of PromptReps with different dense and sparse fusion weights.}
	\label{fig:different_weights}
\end{figure*}

In Figure~\ref{fig:different_weights} we plot the MRR@10 scores obtained with different weights for PromptReps with Llama-8B-Instruct on the MSMARCO dev set. As the plot suggests, the best effectiveness is achieved with weight set to 0.4, noting that 0.5 (the setting used in our zero-shot experiments) actually provides a fairly good choice for hybrid retrieval.

\section{Full results on TREC deep learning and MSMARCO}\label{appendix:results}
In Table~\ref{tab:ablation} we present the full results we abstained on TREC deep learning datasets and MSMARCO passage retrieval dataset. The prompt ID is refer to Table~\ref{tab:prompts}.

\begin{table*}
	\small
	\caption{TREC deep learning and MSMARCO performance of different prompts and LLMs. +BM25 is the system that hybrid dense, sparse, and BM25.}\label{tab:ablation}
	
		\begin{tabular}{r|l|ll|ll|ll}\hline
			Prompt &  &\multicolumn{2}{c|}{DL2019} &\multicolumn{2}{c|}{DL2020} &\multicolumn{2}{c}{MSMARCO Dev} \\
			ID&Methods &nDCG@10 &Recall@1000 &nDCG@10 &Recall@1000 &MRR@10 &Recall@1000 \\\hline
			- &BM25 &49.73 &74.50 &48.76 &80.31 &18.75 &85.73 \\
			- &LLM2Vec &- &- &- &- &13.61 &94.70 \\\hline\hline
			\multicolumn{8}{c}{Phi-3-mini-4k-instruct (3.8B)} \\\hline
			1 &Dense &46.78 &70.10 &42.84 &67.60 &15.45 &82.68 \\
			2 &Dense &34.64 &55.15 &30.62 &50.47 &10.85 &66.04 \\
			3 &Dense &\textbf{49.62} &\textbf{75.79} &\textbf{43.21} &\textbf{71.87} &\textbf{15.78} &\textbf{86.24} \\
			4 &Dense &39.12 &60.77 &28.33 &57.20 &9.26 &72.31 \\
			5 &Dense &43.94 &72.51 &39.00 &70.57 &13.50 &83.08 \\
			6 &Dense &40.77 &62.05 &37.20 &58.39 &14.64 &79.29 \\\hline
			1 &Sparse &41.51 &69.56 &40.95 &69.70 &16.89 &84.72 \\
			2 &Sparse &40.67 &60.59 &39.36 &61.58 &16.38 &75.43 \\
			3 &Sparse &\textbf{42.28} &\textbf{74.33} &40.72 &\textbf{72.14} &18.16 &\textbf{87.09} \\
			4 &Sparse &38.05 &65.15 &34.33 &64.13 &14.75 &78.59 \\
			5 &Sparse &40.68 &70.84 &39.26 &69.02 &16.04 &84.41 \\
			6 &Sparse &41.98 &71.20 &\textbf{41.99} &69.66 &\textbf{18.19} &86.55 \\\hline
			1 &Hybrid &53.04 &79.99 &\textbf{52.76} &77.64 &21.61 &92.21 \\
			2 &Hybrid &50.51 &69.75 &43.92 &66.47 &19.22 &81.35 \\
			3 &Hybrid &\textbf{55.53} &\textbf{81.68} &51.35 &\textbf{79.49} &21.76 &\textbf{93.53} \\
			4 &Hybrid &48.53 &76.29 &40.37 &73.64 &18.23 &87.18 \\
			5 &Hybrid &52.08 &80.16 &50.52 &79.30 &20.30 &92.37 \\
			6 &Hybrid &51.10 &75.84 &49.24 &73.98 &\textbf{22.06} &91.41 \\\hline\hline
			\multicolumn{8}{c}{Meta-Llama-3-8B-Instruct} \\\hline
			1 &Dense &49.26 &\textbf{73.03} &40.28 &68.77 &16.26 &81.96 \\
			2 &Dense &43.32 &64.77 &31.60 &61.35 &12.52 &73.89 \\
			3 &Dense &49.20 &71.69 &\textbf{43.90} &\textbf{69.96} &17.49 &\textbf{84.50} \\
			4 &Dense &0.00 &0.00 &0.00 &0.00 &0.00 &0.04 \\
			5 &Dense &47.19 &72.00 &40.17 &66.71 &16.02 &82.56 \\
			6 &Dense &\textbf{50.62} &73.01 &43.81 &68.39 &\textbf{17.54} &82.91 \\\hline
			1 &Sparse &41.77 &\textbf{67.28} &44.81 &71.36 &20.12 &\textbf{85.71} \\
			2 &Sparse &39.90 &66.00 &43.10 &69.08 &19.13 &83.74 \\
			3 &Sparse &\textbf{43.50} &66.74 &44.87 &\textbf{72.93} &20.42 &85.14 \\
			4 &Sparse &21.77 &41.94 &20.49 &50.51 &7.22 &56.35 \\
			5 &Sparse &42.18 &67.18 &44.17 &71.94 &19.78 &85.37 \\
			6 &Sparse &42.25 &66.58 &\textbf{45.60} &72.82 &\textbf{20.85} &85.57 \\\hline
			1 &Hybrid &53.67 &83.52 &54.35 &78.42 &23.68 &92.84 \\
			2 &Hybrid &50.65 &80.31 &49.25 &76.64 &21.76 &90.12 \\
			3 &Hybrid &55.64 &81.90 &53.83 &79.15 &23.86 &92.99 \\
			4 &Hybrid &13.47 &37.81 &11.81 &45.22 &5.06 &50.50 \\
			5 &Hybrid &54.16 &82.06 &52.06 &78.70 &23.25 &92.77 \\
			6 &Hybrid &55.58 &83.44 &56.66 &79.14 &24.62 &93.11 \\
			6 &+ BM25 &\textbf{63.09} &\textbf{83.82} &\textbf{60.61} &\textbf{79.57} &\textbf{26.75} &\textbf{95.33} \\
			\hline\hline
			\multicolumn{8}{c}{Meta-Llama-3-8B} \\\hline
			6 &Dense &43.90 &67.38 &35.50 &63.34 &14.67 &79.61 \\
			6 &Sparse &38.41 &64.83 &43.34 &67.57 &18.82 &82.63 \\
			6 &Hybrid &\textbf{51.13} &\textbf{77.07} &\textbf{46.34} &\textbf{75.42} &\textbf{22.31} &\textbf{90.87} \\\hline\hline
			\multicolumn{8}{c}{Mistral-7B-Instruct-v0.2} \\\hline
			6 &Dense &13.96 &27.26 &16.77 &26.69 &5.61 &40.27 \\
			6 &Sparse &\textbf{39.84} &\textbf{58.05} &\textbf{37.29} &\textbf{63.53} &\textbf{15.62} &77.55 \\
			6 &Hybrid &32.58 &57.00 &32.95 &63.12 &13.18 &\textbf{77.98} \\\hline\hline
			\multicolumn{8}{c}{Meta-Llama-3-70B-Instruct} \\\hline
			6 &Dense &51.95 &77.30 &45.01 &73.66 &17.76 &85.65 \\
			6 &Sparse &44.07 &68.60 &44.14 &70.99 &20.70 &86.42 \\
			6 &Hybrid &58.39 &86.22 &59.17 &81.57 &25.66 &93.75 \\
			6 &+ BM25 &\textbf{63.18} &\textbf{88.56} &\textbf{62.55} &\textbf{86.28} &\textbf{27.63} &\textbf{95.83} \\\hline
		\end{tabular}

\end{table*}

\section{Fine-tuning hyper-parameters}\label{appendix:parameters}
In Table~\ref{tab:config} we report the fine-tuning hyper-parameters we used for both RepLLama and PromptReps in Section~\ref{sec:supervised}. We use the training data with hard negatives provided in Tevatron Huggingface hub.\footnote{\url{https://huggingface.co/datasets/Tevatron/msmarco-passage-aug}}  

\begin{table*}[t]
	\centering
	\small
	\caption{Hyper-parameters for supervised fine-tuning on MSMARCO passage ranking dataset.}
		\begin{tabular}{l | c}

			\hline
			LLM & LLama3-8B-Instruct \\
			learning rate  & 1e-4 \\
			warmup ratio & 0.1 \\
			per GPU batch size & 8 \\
			\# of GPUs & 4 \\
			gradient accumulation steps  & 4 \\
			\# of negative per example & 15 \\
			total in batch negative & 511 \\
			distance method & cosine similarity \\
			score temperature & 0.01 \\
			query length & 32 \\
			passage length & 156 \\
			LoRA rank & 8 \\
						\hline
		\end{tabular}
	\label{tab:config}
\end{table*}

\section{Full results of different representation methods}\label{appendix:rep_results}
In Table~\ref{tab:full_rep_results} we present the full results of different representation and scoring methods discussed in Section~\ref{sec:reps}. 

\begin{table*}[t]
	\centering
	\caption{Full results of different representation and scoring methods on BEIR.}\label{tab:full_rep_results}
	\resizebox{1\textwidth}{!}{
    \begin{tabular}{l|ccc|ccc|ccc|ccc|ccc}
	\hline
	\multirow{2}{*}{\textbf{Dataset}} & \multicolumn{3}{c}{\textbf{First token single rep}} & \multicolumn{3}{c}{\textbf{First-word single rep}} & \multicolumn{3}{c}{\textbf{Multi token single rep}} & \multicolumn{3}{c}{\textbf{Multi-token multi-rep}} & \multicolumn{3}{c}{\textbf{Multi-word multi-rep}} \\
	& \textbf{Dense} & \textbf{Sparse} & \textbf{Hybrid} & \textbf{Dense} & \textbf{Sparse} & \textbf{Hybrid} & \textbf{Dense} & \textbf{Sparse} & \textbf{Hybrid} & \textbf{Dense} & \textbf{Sparse} & \textbf{Hybrid} & \textbf{Dense} & \textbf{Sparse} & \textbf{Hybrid} \\
	\hline
	{arguana} & 29.70 & 22.85 & 33.32 & 20.54 & 24.59 & 23.80 & 41.78 & 24.46 & 42.61 & 36.69 & 23.03 & 35.19 & 36.47 & 24.13 & 34.96 \\
	{climatefever} & 19.92 & 9.98 & 21.38 & 13.88 & 11.28 & 16.67 & 22.19 & 9.29 & 20.90 & 19.40 & 6.72 & 17.56 & 18.75 & 8.10 & 18.09 \\
	{dbpedia} & 31.53 & 28.84 & 37.71 & 22.71 & 28.70 & 30.08 & 31.83 & 26.33 & 36.03 & 27.46 & 18.18 & 31.35 & 24.50 & 21.66 & 30.32 \\
	{fever} & 56.28 & 52.35 & 71.11 & 40.97 & 57.10 & 61.20 & 50.49 & 51.36 & 64.13 & 44.53 & 30.31 & 54.04 & 38.81 & 37.95 & 52.97 \\
	{fiqa} & 27.11 & 20.33 & 32.40 & 17.61 & 19.60 & 24.74 & 28.94 & 20.73 & 32.44 & 25.26 & 19.41 & 28.38 & 26.28 & 19.50 & 28.30 \\
	{hotpotqa} & 19.64 & 44.75 & 47.05 & 10.35 & 46.25 & 37.79 & 29.94 & 46.50 & 51.50 & 23.80 & 39.39 & 46.38 & 21.93 & 40.25 & 43.68 \\
	{nfcorpus} & 29.56 & 28.18 & 32.98 & 21.98 & 29.15 & 29.49 & 28.97 & 28.65 & 33.65 & 25.68 & 25.39 & 31.20 & 22.95 & 25.37 & 30.05 \\
	{nq} & 34.43 & 29.55 & 43.14 & 22.83 & 29.25 & 33.18 & 35.09 & 25.88 & 39.36 & 31.36 & 23.28 & 35.38 & 30.55 & 22.78 & 35.95 \\
	{quora} & 81.77 & 70.35 & 84.24 & 68.54 & 69.67 & 78.15 & 82.11 & 68.38 & 83.91 & 77.89 & 63.95 & 80.26 & 77.13 & 64.25 & 80.76 \\
	{scidocs} & 18.51 & 11.57 & 17.59 & 12.73 & 12.05 & 14.97 & 18.12 & 11.83 & 17.39 & 16.13 & 11.08 & 15.68 & 15.85 & 11.40 & 15.44 \\
	{scifact} & 52.68 & 58.48 & 65.71 & 26.66 & 58.75 & 51.59 & 52.55 & 59.32 & 63.75 & 45.53 & 54.23 & 58.53 & 47.05 & 53.21 & 61.18 \\
	{trec-covid} & 59.52 & 54.59 & 69.25 & 51.00 & 55.04 & 63.53 & 63.28 & 51.73 & 69.16 & 60.97 & 49.88 & 63.54 & 61.17 & 46.24 & 65.10 \\
	{touche2020} & 14.85 & 18.47 & 21.65 & 12.23 & 21.58 & 19.13 & 15.59 & 19.24 & 21.44 & 15.86 & 17.81 & 22.10 & 15.41 & 18.09 & 19.26 \\\hline
	{avg} & 36.58 & 34.64 & 44.43 & 26.31 & 35.62 & 37.26 & 38.53 & 34.13 & 44.33 & 34.65 & 29.44 & 39.97 & 33.60 & 30.23 & 39.70 \\
	\hline
\end{tabular}
	}
\end{table*}




\end{document}